\documentclass[a4paper,11pt]{article}
\pdfoutput=1 

\usepackage{jheppub}
\usepackage{amsmath}
\usepackage{amssymb}
\usepackage{epsfig}
\usepackage{subfigure}
\usepackage{cancel}
\usepackage{paralist}

\usepackage[normalem]{ulem}

\newcommand{\at}{\makeatletter @\makeatother}

\newcommand{\GeV}{ {\rm GeV}}

\def\msbar{\ensuremath{{\rm{\overline{MS}}}}}

\newcommand{\der}{\ensuremath{{\operatorname{d}}}}

\title{On the intrinsic bottom content of the nucleon and its impact on heavy new physics at the LHC}

\author[a]{Florian Lyonnet,}
\author[b]{Aleksander Kusina,}
\author[c]{Tom\'{a}\v{s} Je\v{z}o,}
\author[d]{Karol Kova\v{r}\'{\i}k,}
\author[a]{Fred Olness,}
\author[b]{Ingo\ Schienbein,}
\author[a]{Ji-Young\ Yu}

\affiliation[a]{Southern Methodist University,Dallas, TX 75275, USA}

\affiliation[b]{
Laboratoire de Physique Subatomique et de Cosmologie, Universit\'e Grenoble-Alpes, CNRS/IN2P3,
\\
53 avenue des Martyrs, 38026 Grenoble, France}
\affiliation[c]{Universit\`a di Milano-Bicocca and INFN, Sezione di Milano-Bicocca,\\
Piazza della Scienza 3, 20126 Milano, Italy}

\affiliation[d]{Institut f{\"u}r Theoretische Physik, Westf{\"a}lische Wilhelms-Universit{\"a}t M{\"u}nster,
             Wilhelm-Klemm-Stra{\ss}e 9, D-48149 M{\"u}nster, Germany}

\emailAdd{flyonnet@mail.smu.edu}
\emailAdd{akusina@mail.smu.edu}
\emailAdd{tomas.jezo@mib.infn.it}
\emailAdd{kovarik@particle.uni-karlsruhe.de}
\emailAdd{olness@physics.smu.edu}
\emailAdd{schien@lpsc.in2p3.fr}
\emailAdd{yu@physics.smu.edu}

\abstract{

Heavy quark parton distribution functions (PDFs) play an important role
in several Standard Model and New Physics processes. 
Most analyses rely on the assumption that the charm and bottom PDFs are
generated perturbatively by gluon splitting and do not involve any non-perturbative
degrees of freedom. It is clearly necessary to test this hypothesis
with suitable QCD processes.
Conversely, a non-perturbative, intrinsic heavy quark parton distribution
has been predicted in the literature. 
We demonstrate that to a very good approximation the scale-evolution
of the intrinsic heavy quark content of the nucleon is governed by
non-singlet evolution equations. This allows us to analyze the intrinsic heavy
quark distributions without having to resort to a full-fledged global analysis
of parton distribution functions. We exploit this freedom to model intrinsic
bottom distributions which are so far missing in the literature in order to 
estimate the impact of this non-perturbative contribution to the bottom-quark
PDF, and on parton--parton luminosities at the LHC.
This technique can be applied to the case of intrinsic charm, 
albeit within the limitations outlined in the following. 
}


\preprint{LPSC-15-083, SMU-HEP-15-01, MS-TP-15-06}

\begin{document} 
\maketitle

\section{Introduction}

Heavy quark parton distribution functions (PDFs) play an important role
in several Standard Model (SM) and New Physics (NP) processes at the
CERN Large Hadron Collider (LHC). 
In particular, several key processes involve the bottom quark
PDF, e.g. $tW$, $tH^+$ production, associated $b$ plus $W/Z/H$ boson production
or $Hbb$ production~\cite{Maltoni:2012pa}.
In the standard approach employed by almost all global analyses of PDFs,
the heavy quark distributions  are generated {\em radiatively}, according
to DGLAP evolution equations~\cite{Altarelli:1977zs,Gribov:1972ri,Dokshitzer:1977sg}, 
starting with a perturbatively calculable boundary condition~\cite{Collins:1986mp,Buza:1996wv} 
at a scale of the order of the heavy quark mass.%
    \footnote{The most common approach is to use one of the general mass
    variable flavor number schemes (GM VFNS) such as ACOT~\cite{Aivazis:1993pi,Collins:1998rz}
    or FONLL~\cite{Forte:2010ta,Cacciari:1998it}.
    For a review of the treatment of heavy quarks in PDF global analyses
    see e.g.~\cite{Thorne:1998xv}.}
In other words, there are no free fit parameters associated with the heavy
quark distribution and it is entirely related to the gluon PDF at the scale
of the boundary condition.
As a consequence, the uncertainties for the heavy quark and gluon
distributions are strongly correlated; this has been discussed in the context
of inclusive Higgs production at the Tevatron and LHC~\cite{Belyaev:2005nu}.

	However, a purely perturbative, {\it extrinsic}, treatment where the heavy quarks are 
radiatively generated might {\em not} be adequate;
in particular for the charm quark with a mass $m_c \simeq 1.3$ GeV which is
not much bigger than typical hadronic scales but also for the bottom quark
with a mass $m_b \simeq 4.5$ GeV.
Indeed, there are a number of models that postulate a non-perturbative, {\it intrinsic},  heavy
quark component which is present even for scales $Q$ below the heavy quark mass $m$.
%
In particular,
light-cone models predict a non-perturbative ('intrinsic') heavy quark
component in the proton wave-function~\cite{Brodsky:1980pb,Brodsky:1981se}
and similar expectations result from meson cloud
models~\cite{Navarra:1995rq,Paiva:1996dd,Melnitchouk:1997ig};
an overview of different models can be found, e.g., in \cite{Pumplin:2005yf}. 
Predictions of these models together with the EMC charm data~\cite{Aubert:1982tt,Aubert:1981ix}
motivated first theoretical analyses of the intrinsic charm (IC) content of the
proton~\cite{Hoffmann:1983ah,Harris:1995jx,Steffens:1999hx}. These first analyses
were not global as they concentrated only on the possibility of explaining the
EMC data. Later the CTEQ collaboration performed the first fully global analyses
of PDFs including the IC possibility~\cite{Pumplin:2007wg,Nadolsky:2008zw}.
These studies gave the first estimate, based on an array of different data sets,
of how big the intrinsic charm could be. The possibility of IC was  also
considered by the MSTW group~\cite{Martin:2009iq}. 
Most recently, two new global PDF analyses dedicated to  IC have been performed:%
    \footnote{From private communication we also know that the NNPDF collaboration
    will soon release an IC analysis based on the NNPDF3.0 framework~\cite{Ball:2014uwa}.}
(i) the  CTEQ collaboration has updated  their previous work~\cite{Dulat:2013hea}
using the CT10 NNLO framework~\cite{Gao:2013xoa},
and (ii) an analysis of Jimenez-Delgado {\em et al.}~\cite{Jimenez-Delgado:2014zga}
which is interesting as it uses less strict kinematic cuts which allow for
the inclusion of low-$Q$, high-$x$ data that should be more sensitive to the
light-cone inspired IC component.%
    \footnote{
    Of course this requires including additional corrections (target
    mass and higher twist) as the leading twist approximation does not necessarily
    hold in this kinematic regime. Also, as some of the data used is on heavy
    nuclear targets, nuclear corrections were employed as well. For an instructive discussion
    concerning this work see~\cite{Brodsky:2015uwa}.}
These two most recent analyses set significantly different limits on the allowed
IC contribution, partly because of the very different tolerance criteria which are used to
define the range of acceptable fits.
These differences highlight the utility of the techniques discussed in this 
paper as we can freely adjust the amount of IC/IB contributions 
(within limits, which we will quantify)
 without having to regenerate a complete global analysis for each case. 

It is essential to experimentally test the  heavy quark PDFs, both the extrinsic and intrinsic components.
One observable which is directly sensitive to an IC component
is the deep inelastic charm structure function $F_2^c(x,Q^2)$.
So far, the EMC data for $F_2^c$ is the only measurement of 
the charm structure function in the relevant $(x,Q^2)$ region which is 
sensitive to a large-$x$ IC component; this is the only DIS data
cited as evidence for intrinsic charm.
The HERA data on $F_2^c$ and $F_2^b$ exist only for $x < 0.1$ and provide no constraints 
on a large-$x$ IC or IB; only the inclusive structure function $F_2(x,Q^2)$ at large-$x$
measured quite precisely at HERA provides limited information on the large-$x$ charm PDF.
At a future Electron-Ion Collider (EIC) the large-$x$ charm structure function would be accessible, 
and the rate for charm production at $x \gtrsim 0.1$
could be increased by up to an order of magnitude due to the presence of a large-$x$ non-perturbative
intrinsic charm component in the nucleon \cite{Boer:2011fh}. 
Alternatively, the IC could be searched for by measuring the 
Callan-Gross ratio $R(x,Q^2) = F_L/F_T$ \cite{Ivanov:2008xu}, 
or by studying angular distributions  \cite{Ananikyan:2007ef,Ananikyan:2006kz}.

In hadronic collisions, a promising way to constrain models of IC is the
measurement of inclusive charm hadron production ($D^0, D^+, D^{\star +}, \Lambda_c, \ldots$).
Predictions for such processes were obtained in the general-mass variable-flavor-number scheme (GM-VFNS) 
\cite{Kniehl:2004fy,Kniehl:2005mk,Kniehl:2005st} for the Tevatron at a center-of-mass (cms) energy of 1960 GeV, for
the Relativistic Heavy Ion Collider (RHIC)  at cms energies of 200 GeV (RHIC200) and 500 GeV (RHIC500), 
and for the LHC at a cms energy of 7 TeV (LHC7) \cite{Spiesberger:2012zn,Kniehl:2012ti,Kniehl:2011bk,Kniehl:2009ar}.
The IC charm effects can be particularly large at the RHIC200 and at the LHC7 at
forward rapidities where the differential cross section can be enhanced by a
factor of up to 5 compared to the prediction with a radiatively generated charm PDF \cite{Kniehl:2012ti}.
Recent results from LHCb \cite{Aaij:2013mga} can serve as an example of such data
which are, however, not yet precise enough and do not extend to sufficiently large transverse
momenta at the most forward rapidities ($4 <  y < 4.5$) to be conclusive.
Another process which is very sensitive to the heavy quark PDF is direct 
photon production in association with a heavy quark jet \cite{Kovarik:2012pj}.
Data from the D0 experiment at the Tevatron \cite{Abazov:2009de,Abazov:2012ea}
overshoot the standard NLO QCD predictions \cite{Stavreva:2009vi} at large
transverse photon momenta; the inclusion of an intrinsic heavy quark
component in the nucleon can reduce the difference between data and theory, 
but not fully resolve it.
In fact, at the Tevatron the $q \bar q$ channel becomes important at large transverse momenta, 
and higher-order corrections to this channel not included in Ref.~\cite{Stavreva:2009vi} might
explain (part of) this discrepancy.
The $q \bar q$-channel does not play an important role at $pp$ colliders; 
therefore, measurements of this process at RHIC and the LHC probe the heavy quark PDFs
in different regions of the momentum fraction $x$ and could
shed more light on the current situation.\footnote{In addition, the measurements at the LHC would provide an important baseline for
$\gamma+Q$ production in $pA$~\cite{Stavreva:2010mw}
and $AA$ collisions~\cite{Stavreva:2011wc}.}
A detailed study of $\gamma + Q$ production at the LHC operating at $\sqrt{S}=8$ TeV (LHC8) 
was performed in Refs.~\cite{Bednyakov:2014pqa,Bednyakov:2013zta}.
There it was shown that the existence of IC in the proton can be visible at large transverse momenta
of the photons and heavy quark jets at rapidities $1.5 < |y_\gamma|<2.4, |y_c|<2.4$.
Indeed, for the BHPS model~\cite{Brodsky:1980pb} the cross section can be enhanced by a factor of 2 or 3 for $p_T^\gamma > 200$~GeV.
However, the cross section is already quite small in this kinematic range so that this measurement will be statistically limited.
The ideal place to observe or constrain intrinsic charm would be {\it A Fixed Target ExpeRiment} using the LHC beams
(AFTER\at LHC) 
\cite{Brodsky:2012vg,Lansberg:2012kf,Lansberg:2013wpx,Rakotozafindrabe:2013cmt} 
due to the lower cms energy ($\sqrt{S}=115$ GeV) together with a very high luminosity;
for a review see \cite{review_ic}.

While there are at least a few global analyses 
which allow for  an intrinsic charm component in the nucleon
\cite{Pumplin:2007wg,Nadolsky:2008zw,Dulat:2013hea,Jimenez-Delgado:2014zga},
studies of intrinsic bottom PDFs have not been performed at all.
The main purpose of this paper is to 
outline a technique which can provide IB PDFs 
for any generic non-IB PDF set; 
we can then directly compare the  IB PDFs  with the  non-IB PDFs 
to  gauge the impact of the  non-perturbative IB component of the nucleon structure on
$b$-quark initiated processes. 
Our approach exploits the fact that  the intrinsic bottom PDF evolves (to an excellent precision)
according to a standalone non-singlet evolution equation.
Furthermore, due to the small momentum fraction carried by the IB PDF,
the evolution of the other partons is essentially not disturbed by the IB component.
These two observation allows us to compute the IB PDF  {\it without} the need to
perform a complete global analysis of PDFs. 
Thus, we can easily obtain a matched set of IB and non-IB PDFs. 
Note that because  existing data entering global analyses of proton PDFs do not constrain 
the IB PDF, it would not be useful to try and 
obtain information on the IB content of the nucleon using a global fit. 


The rest of this paper is organized as follows.
In Sec.~\ref{sec:intrinsic}, we demonstrate that the scale-evolution
of the intrinsic PDF is governed by a non-singlet evolution equation. We then
propose  suitable boundary conditions and perform a number of numerical
tests of the quality of our approximations. In particular, we investigate to what
degree the gluon distribution is perturbed by the presence of the intrinsic  component.
In Sec.~\ref{sec:numerics}, we use the IB PDFs to obtain predictions
for parton--parton luminosities relevant at the LHC.
In Sec.~\ref{sec:discussion}, we discuss our results
and assess the impact of the intrinsic  component. 
Finally, in Sec.~\ref{sec:conclusions}, we summarize our results and
present conclusions.

\section{Intrinsic heavy quark PDFs} 
\label{sec:intrinsic}

\subsection{Definition}
\label{sec:definition}
In the context of a global analysis of PDFs the different
parton flavors are specified via a boundary condition at
the input scale $\mu_0$ which is typically of the order
${\cal O}(1\ \GeV)$. Solving the DGLAP evolution equations
with these boundary conditions allows us to determine the PDFs
at higher scales $\mu > \mu_0$.
The boundary conditions for the up, down, strange quarks and
gluons are not perturbatively calculable and have to be determined
from experimental data. From this perspective, it is meaningless to decompose the light quark and gluons PDFs
into distinct (extrinsic and intrinsic) components.
The situation is different for the heavy charm and bottom quarks; 
the boundary conditions have been calculated perturbatively and
resum to all orders collinear logarithms associated 
with the heavy quark lines at fixed-order  in perturbation theory.
A non-perturbative (intrinsic) heavy quark distribution $Q_1$ can then be
defined at the input scale $\mu_0$ as the difference of the full boundary condition for the
heavy quark PDF $Q$
and the perturbatively calculable (extrinsic) boundary condition $Q_0$:
\begin{equation}
Q_1(x,\mu_0):= Q(x,\mu_0) - Q_0(x,\mu_0)\, ,
\label{eq:Q1}
\end{equation}
where $Q=c$ or $Q=b$.
At NLO in the $\msbar$ scheme, the relation in Eq.~\eqref{eq:Q1} gets further simplified
if the input scale $\mu_0$ is identified with the heavy quark mass $m_Q$ because $Q_0(x,m_Q)=0$ 
at NLO.
In this case, any non-zero boundary condition $Q(x,m_Q) \ne 0$ can be attributed to the intrinsic
heavy quark component.
This simplification, however, does not hold a priori for a different factorization scheme, nor at NNLO and beyond.


\subsection{Evolution}
\label{sec:evolution}

In the following, we demonstrate that to a good approximation the intrinsic heavy quark distributions
are governed by non-singlet evolution equations. Denoting 
the vector of light quarks as `$q$' and the
heavy quark
distribution by `$Q$' (where $Q=c$ or $Q=b$) the 
Dokshitzer-Gribov-Lipatov-Altarelli-Parisi (DGLAP) evolution equations read
\cite{Altarelli:1977zs,Gribov:1972ri,Dokshitzer:1977sg}
\begin{eqnarray}
\label{eq:DGLAP1a}
\dot g &=& P_{gg}\otimes g+P_{gq}\otimes q+P_{gQ}\otimes Q \, ,
\\
\label{eq:DGLAP1b}
\dot q &=& P_{qg}\otimes g+P_{qq}\otimes q+P_{qQ}\otimes Q \, ,
\\
\label{eq:DGLAP1c}
\dot Q &=& P_{Qg}\otimes g+P_{Qq}\otimes q+P_{QQ}\otimes Q \, ,
\end{eqnarray}
with the splitting functions $P_{Qg}(x)=P_{qg}(x)$, $P_{QQ}(x)=P_{qq}(x)$, $P_{Qq}(x) = P_{q'q}(x)$, etc.
in the massless $\msbar$ scheme which are known up to three-loop order \cite{Vogt:2004mw,Moch:2004pa}. 

Next we substitute $Q=Q_0+Q_1$ where $Q_0$ denotes the usual radiatively generated
extrinsic heavy quark component and $Q_1$ is the non-perturbative intrinsic heavy quark
distribution:\footnote{Strictly speaking, the decomposition of $Q$ into $Q_0$ and $Q_1$ is 
defined at the input scale where the calculable boundary condition for $Q_0$ is known. 
Consequently, $Q_1:=Q-Q_0$ is known as well.
Only due to the approximations in Eqs.\ \protect\eqref{eq:DGLAP2a} and \protect\eqref{eq:DGLAP2b}
it is possible to entirely decouple $Q_0$ from $Q_1$ so that the decomposition becomes meaningful
at any scale.} 
\begin{eqnarray}
\label{eq:DGLAP2a}
\dot g &=& P_{gg}\otimes g+P_{gq}\otimes q+P_{gQ}\otimes Q_0+ {\cancel{P_{gQ}\otimes Q_1}}\, ,
 \\
\label{eq:DGLAP2b}
\dot q &=& P_{qg}\otimes g+P_{qq}\otimes q+P_{qQ}\otimes Q_0+ {\cancel{P_{qQ}\otimes Q_1}}\, ,
 \\
\label{eq:DGLAP2c}
\dot Q_0 + \dot Q_1 &=& P_{Qg}\otimes g+P_{Qq}\otimes q+P_{QQ}\otimes Q_0+ P_{QQ}\otimes Q_1 \, .
\end{eqnarray}
Neglecting the crossed out terms which give a tiny contribution 
to the evolution of the gluon and light quark distributions the
system of evolution equations can be separated into two independent parts.
For the system of gluon, light quarks and extrinsic heavy
quark ($g,q,Q_0$) one recovers the same evolution equations as in the standard approach
without an intrinsic heavy quark component
\begin{eqnarray}
\label{eq:DGLAP3a}
\dot g &=& P_{gg}\otimes g+P_{gq}\otimes q+P_{gQ}\otimes Q_0 \, ,
 \\
\label{eq:DGLAP3b}
\dot q &=& P_{qg}\otimes g+P_{qq}\otimes q+P_{qQ}\otimes Q_0 \, ,
 \\
\label{eq:DGLAP3c}
\dot Q_0 &=& P_{Qg}\otimes g+P_{Qq}\otimes q+P_{QQ}\otimes Q_0 \, .
\end{eqnarray}
For the intrinsic heavy quark distribution, $Q_1$, one finds
a standalone non-singlet evolution equation
\begin{equation}
\dot Q_1 = P_{QQ}\otimes Q_1 \, . 
\label{eq:DGLAP4}
\end{equation}

In a global analysis with intrinsic heavy quark PDFs, 
using the exact evolution equations \eqref{eq:DGLAP1a}--\eqref{eq:DGLAP1c},
the parton distributions satisfy the momentum sum rule
\begin{equation}
\int_0^1\ \der x\  x\ \left(g + \sum_i (q_i+\bar{q}_i ) + Q_0 + \bar{Q}_0 +Q_1 + \bar{Q}_1 \right) = 1\, . 
\end{equation}
Allowing for a small violation of this sum rule it is possible to
entirely decouple the analysis of the intrinsic heavy quark distribution
from the rest of the system.
The PDFs for the gluon, the light quarks and the extrinsic heavy quark
can be taken from a global analysis in the standard approach 
using Eqs.\ \eqref{eq:DGLAP3a} -- \eqref{eq:DGLAP3c}
where they already saturate the momentum sum rule
\begin{equation}
\int_0^1\ \der x\  x\ \left(g + \sum_i (q_i+\bar{q}_i ) + Q_0 + \bar{Q}_0 \right) = 1\, . 
\end{equation}
On top of these PDFs the intrinsic heavy quark PDF can be determined
in a standalone analysis using the non-singlet evolution equation \eqref{eq:DGLAP4}.
This induces a violation of the momentum sum rule by the term
\begin{equation}
\int_0^1\ \der x\  x\  \left(Q_1 + \bar{Q}_1\right)  
\end{equation}
which, however, is very small for bottom quarks.%
    \footnote{It is also acceptable in case of charm provided that the
    allowed normalization of IC is not too big.}
We will perform numerical checks of the validity of our approximations in Sec.~\ref{sec:tests}
after having discussed the boundary conditions for the intrinsic heavy quark distribution.

\subsection{Modeling the boundary condition}
\label{sec:bc}

The BHPS model~\cite{Brodsky:1980pb} predicts the following $x$-dependence
for the intrinsic charm (IC) parton distribution function:
\begin{equation}
c_1(x) = \bar c_1(x) \propto x^2 [6 x (1+x) \ln x + (1-x)(1+10 x+x^2)]\, .
\label{eq:bhps}
\end{equation}
Conversely, the normalization and the precise energy scale of this distribution are
not specified.
In the CTEQ global analyses with intrinsic charm~\cite{Pumplin:2007wg,Nadolsky:2008zw} this functional form has been
used as a boundary condition at the scale $Q = m_c$ leaving the normalization 
as a free fit parameter.

We expect the $x$-shape of the intrinsic bottom distribution $b_1(x)$
to be very similar to the one of the intrinsic charm distribution.
%
Furthermore, the normalization of IB is expected to be parametrically 
suppressed with respect to IC by a factor $m_c^2/m_b^2 \simeq 0.1$.
%
Therefore, because the scale of the boundary condition is not fixed, the following two ansatzes for $b_1$ can be considered
\begin{equation}
{\rm \it Different\ Scales: } \qquad
b_1(x,m_b) = \frac{m_c^2}{m_b^2}\, c_1(x,m_c)\, , 
\label{eq:bc1}
\end{equation}
%
\begin{equation}
{\rm \it Same\ Scales: } \qquad
b_1(x,m_c) = \frac{m_c^2}{m_b^2} c_1(x,m_c)\, .
\label{eq:bc2}
\end{equation}
%
In the following we use the {\it Same Scales} boundary condition, Eq.~\eqref{eq:bc2}, which remains valid at any scale $Q$.

In this case, since $c_1 = c - c_0$,  it is possible to construct the IB PDF
from the difference of the CTEQ6.6c and the standard CTEQ6.6 charm PDFs at any scale
without having to solve the non-singlet evolution equation for the IB PDF.
We will compare the two boundary conditions in Eqs.\ \eqref{eq:bc1} and \eqref{eq:bc2}
in Sec.~\ref{sec:tests}.
Finally, let us note that it would be no problem to work with asymmetric boundary conditions, $\bar c_1(x) \ne c_1(x)$
and $\bar b_1(x) \ne b_1(x)$, as predicted for example by meson cloud models~\cite{Paiva:1996dd}.

\subsection{Intrinsic heavy quark PDFs from non-singlet evolution}
\label{sec:PDFdef}

\begin{figure}
\begin{center}
\includegraphics[angle=0,width=0.49\textwidth]{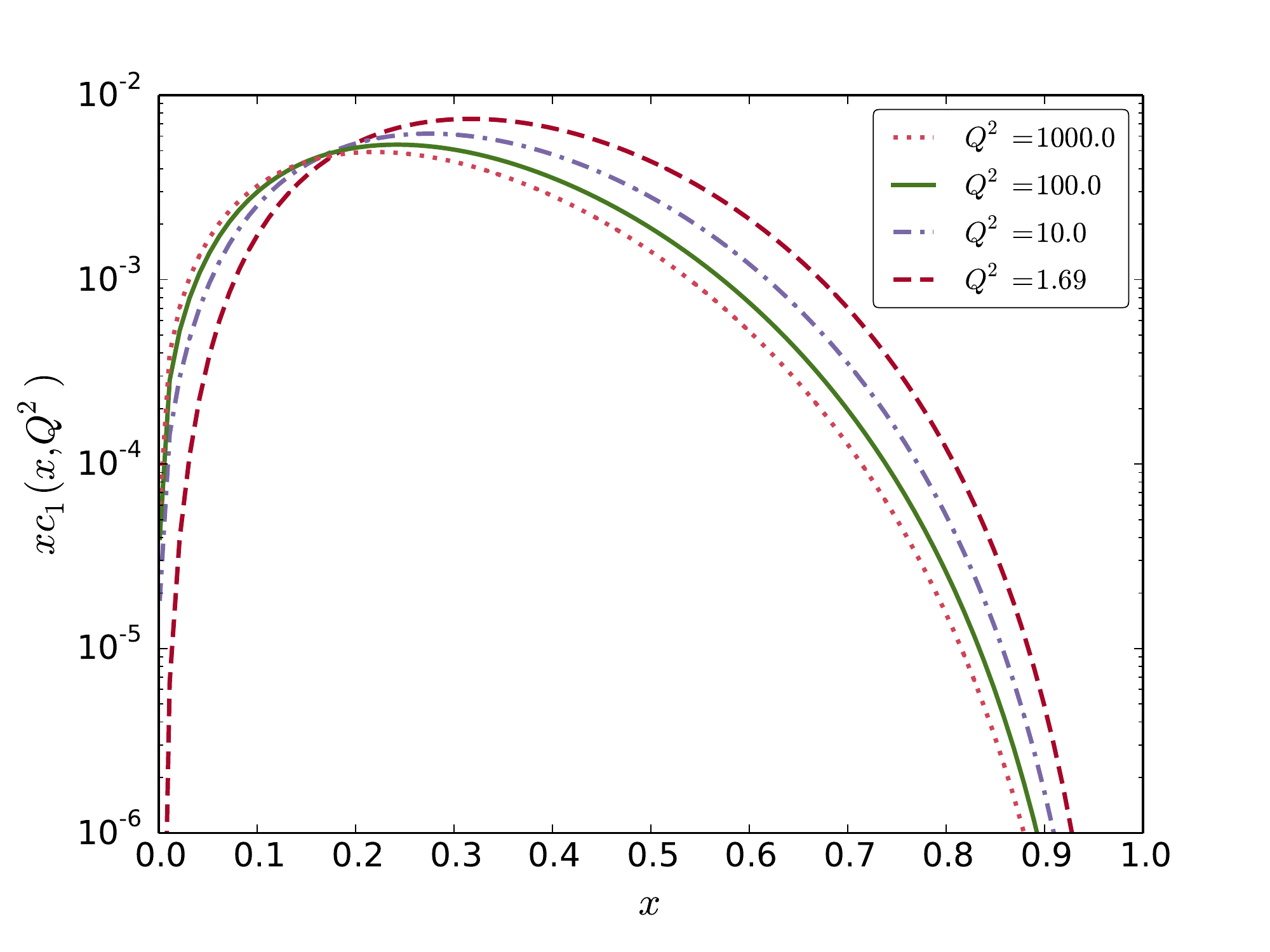}
\includegraphics[angle=0,width=0.49\textwidth]{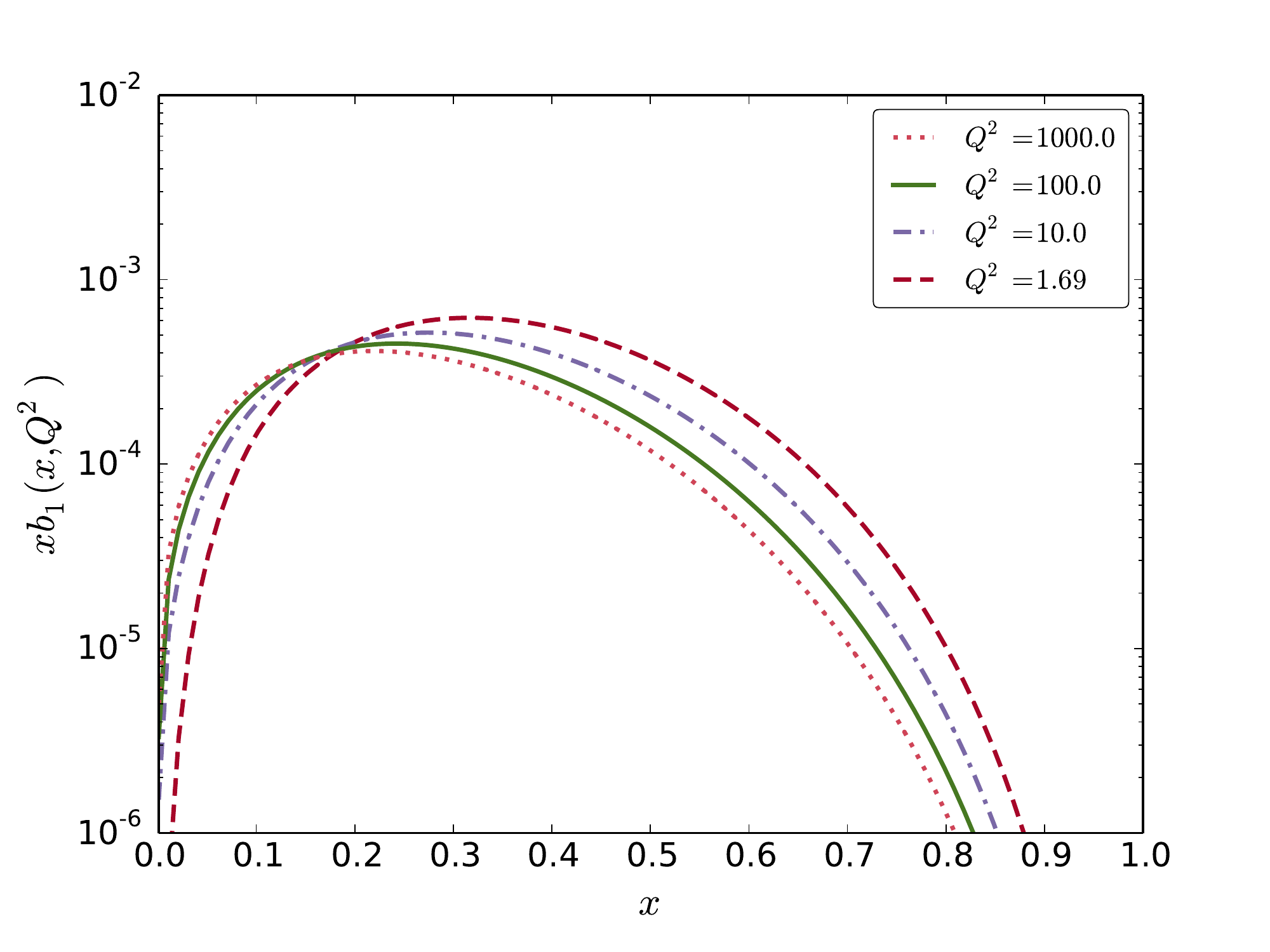}
\caption{Scale-evolution of the IC (left) and IB (right) PDF at NLO according to the non-singlet evolution equation \eqref{eq:DGLAP4} 
using the {\it Same Scales} boundary condition in Eq.~\eqref{eq:bc2}  with $m_c = 1.3$ and $m_b = 4.5$~GeV. 
Shown are results for $Q^2= 1.69, 10, 100, 1000$, and $10000$ GeV$^2$.}
\label{fig:scale-evolution}
\end{center}
\end{figure}

For the purpose of this analysis we used the approximation of
Sec.~\ref{sec:evolution} to 
produce standalone IC and IB PDFs that can be used together
with any regular PDF set sharing the same values for the QCD parameters, such as the strong coupling or the quark masses.
For the IC PDF we used Eq.~\eqref{eq:bhps} to define the initial
$x$-dependence at the scale of the charm mass, and fixed the normalization
to match the one predicted by the CTEQ6.6c0 fit~\cite{Nadolsky:2008zw}.
The IB PDF was generated using the 
{\it Same Scales}
boundary conditions of Eq.~\eqref{eq:bc2}
together with the same $x$-dependent input of Eq.~\eqref{eq:bhps}.
If not stated otherwise, the normalization for the IB PDF was chosen to be 
identical to the IC case scaled down by a factor $m_c^2/m_b^2 = 0.083$.
Both PDFs were then evolved according to the non-singlet evolution
equation~\eqref{eq:DGLAP4}  
and the corresponding grids were produced.%
    \footnote{The evolution was performed in Mellin-moment space using the PEGASUS package~\cite{Vogt:2004ns}
    at NLO in the $\msbar$ scheme. Additionally the strong coupling was
    chosen so that it corresponds to the one used in the CTEQ6.6 set~\cite{Nadolsky:2008zw}.}
We show these distributions for selected values of the factorization scale ($Q$) in
Fig.~\ref{fig:scale-evolution}.
As in our approximation, the evolution of the intrinsic charm and bottom PDFs
is completely decoupled from the light quarks, gluons, and perturbative heavy
quark components; the normalization of our PDFs can be easily changed by
means of simple rescaling.
However, for convenience we also produced a set with normalization
corresponding to the CTEQ6.6c1 fit~\cite{Nadolsky:2008zw} allowing for
a larger intrinsic component.

\subsection{Numerical validation}
\label{sec:tests}


In order to test the ideas presented in Secs.~\ref{sec:evolution} and \ref{sec:bc}, we use the
CTEQ6.6c series of intrinsic charm fits which have been obtained in the framework of
the CTEQ6.6 global analysis \cite{Nadolsky:2008zw}. 
The CTEQ6.6c series comprises 4 sets of PDFs including an intrinsic charm component.
Two of them, CTEQ6.6c0 and CTEQ6.6c1, employ the BHPS model with $1\%$ and $3.5\%$ IC probability, respectively.%
     \footnote{The other two sets, CTEQ6.6c2 and CTEQ6.6c3, study a 'sea-like' intrinsic
     charm with low and high strength, respectively. We don't consider them in this paper
     because they are theoretically less motivated. Note also that the picture of a
     non-singlet intrinsic heavy quark distribution does not naturally apply, since these
     distributions are substantial in the small-$x$ region. Additional numerical tests
     would be needed in this case.}
This corresponds to the values of 0.01 and 0.035 of the first moment of the charm PDF,
$\int dx  \, c(x)$, calculated at the input scale $Q_0=m_c=1.3\ \GeV$. For a review of these
models see Ref.\ \cite{Pumplin:2007wg}.
In the rest of this article, we will follow the naming convention of the CTEQ6.6c fits
in which a given fit is characterized by the value in percentage of the first moment of
the charm distribution at
the input scale, e.g. 1\% for CTEQ6.6c0. 
For convenience, we list below the first and second moments (calculated at the input scale)
for the sets referred to in the following. 
\begin{center}
\begin{tabular}{|l|c|c|}
\hline
          &  $\int_0^1 dx \;c(x)$ & $\int_0^1 dx \;x \left[c(x)+\bar{c}(x)\right] \equiv <x>_{c+\bar{c}}$ \\
\hline
\hline
CTEQ6.6   &  0                    &  0 \\
\hline
CTEQ6.6c0 &  0.01                 &  0.0057 \\
\hline
CTEQ6.6c1 &  0.035                &  0.0200 \\
\hline
\end{tabular}
\end{center}
As can be seen the actual momentum carried by the charm in the CTEQ6.6c0 and CTEQ6.6c1 fits
is equal to $\sim0.6$\% and 2\% respectively.

In the following we compare our approximate IC PDFs supplemented with the central
CTEQ6.6 fit, which has a radiatively generated charm distribution, with the CTEQ6.6c0
and CTEQ6.6c1 sets where IC has been obtained from global analysis without the approximations
of Sec.~\ref{sec:evolution}.

\begin{figure}
\begin{center}
\subfigure[]{
\label{fig:testa}
\includegraphics[angle=0,width=0.48\textwidth]{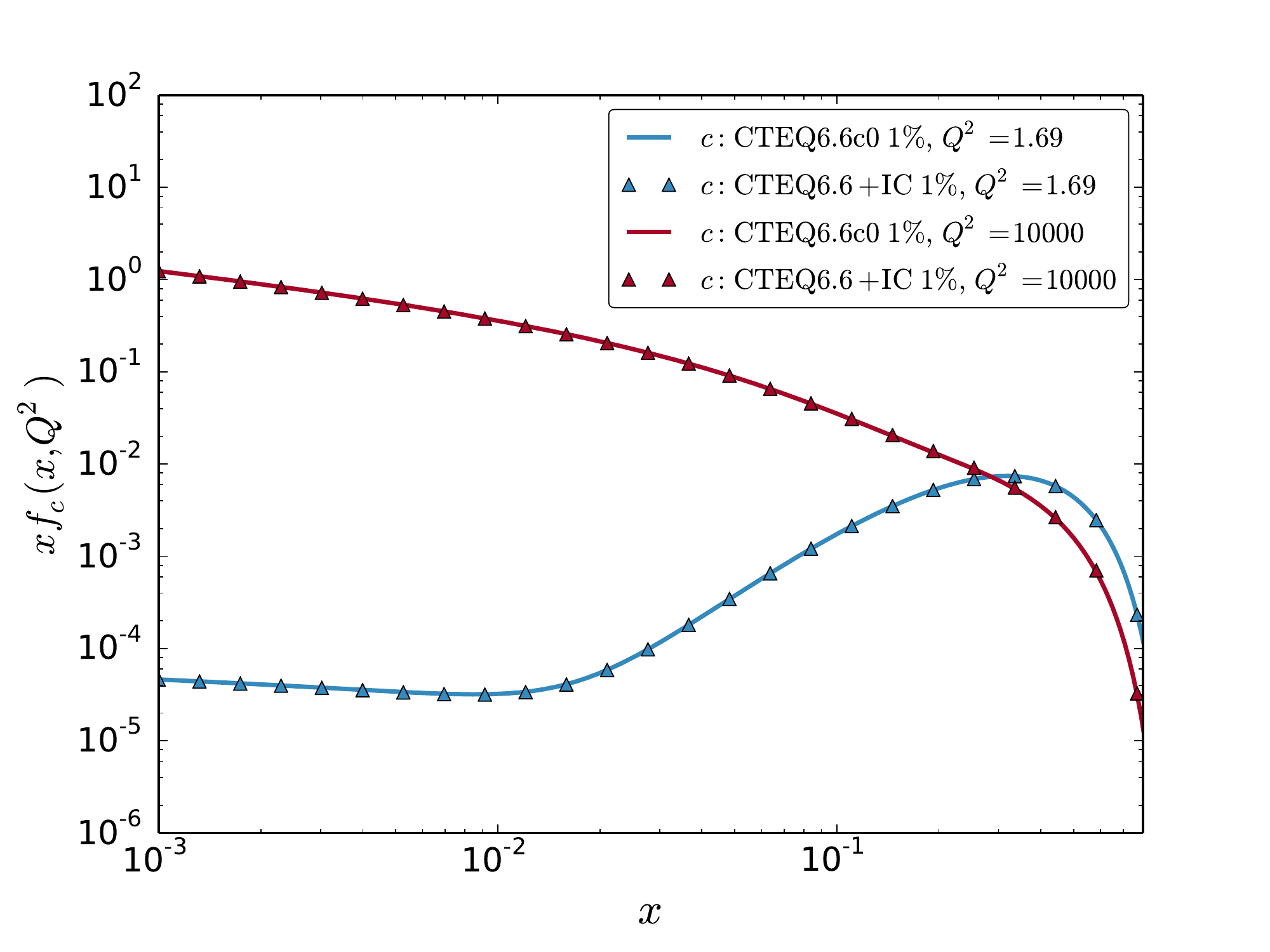}}
\subfigure[]{
\label{fig:testb}
\includegraphics[angle=0,width=0.48\textwidth]{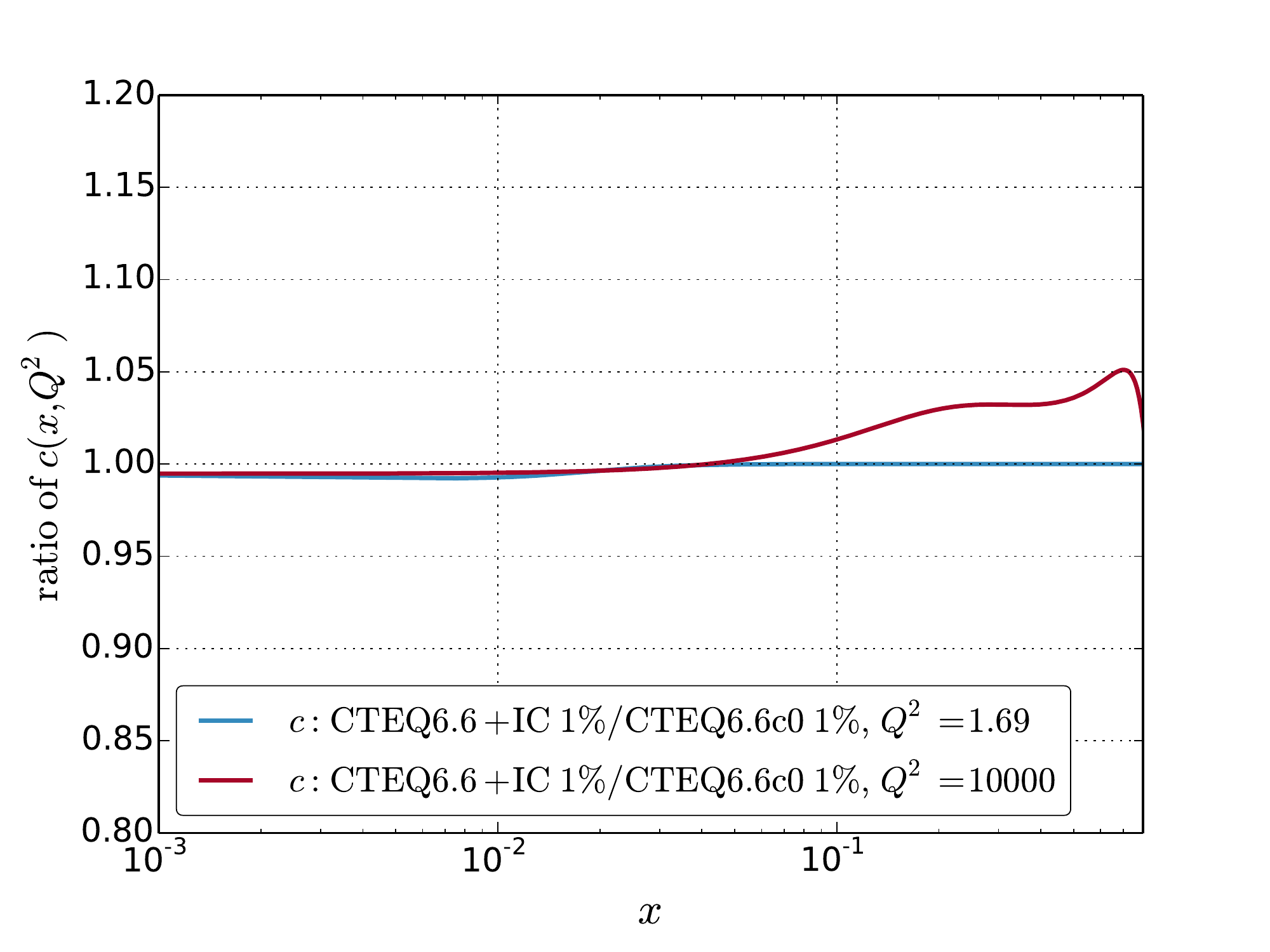}}
\caption{(a) CTEQ6.6c0 charm distribution function $c(x,Q^2)$ (solid lines)
and the sum $c_0(x,Q^2) + c_1(x,Q^2)$ (triangles) where $c_0$ is the radiatively
generated CTEQ6.6 charm distribution and $c_1$
is the non-singlet evolved IC using the BHPS boundary condition \protect\eqref{eq:bhps}
with the same normalization as used for the CTEQ6.6c0 charm distribution.
Results are shown for the input scale $Q^2=Q_0^2=m_c^2$ and the scale $Q^2=10000\ \GeV^2$.
Fig.~(b) shows the ratio of the curves in (a).
}
\label{fig:test}
\end{center}
\end{figure}

In Fig.~\ref{fig:testa} the CTEQ6.6c0 charm distribution function $c(x,Q^2)$ is shown (solid lines) for two scales, $Q^2=1.69$ and $10000\ \GeV^2$, 
in dependence of $x$.
The doted lines have been obtained as the sum of  $c_0(x,Q^2) + c_1(x,Q^2)$ where $c_0$ is the radiatively
generated charm distribution using the CTEQ6.6 PDF and $c_1$
is the non-singlet evolved IC using the boundary condition \eqref{eq:bhps} with the same normalization as used for the 
CTEQ6.6c0 charm distribution.
%
As can be seen in the ratio plot, Fig.~\ref{fig:testb}, 
the difference between the sum $c_0+c_1$ and the CTEQ6.6c0 charm distribution 
is tiny at low $Q^2$, and smaller than $5\%$ at the higher $Q^2$. 
In other words, 
the IC distribution $c_1$ evolved according to the decoupled non-singlet evolution equation
is in very good agreement with 
the difference $c-c_0$ representing the IC component in the full global analysis.
%

\begin{figure}
\begin{center}
\subfigure[]{
\label{fig:gluona}
\includegraphics[angle=0,width=0.48\textwidth]{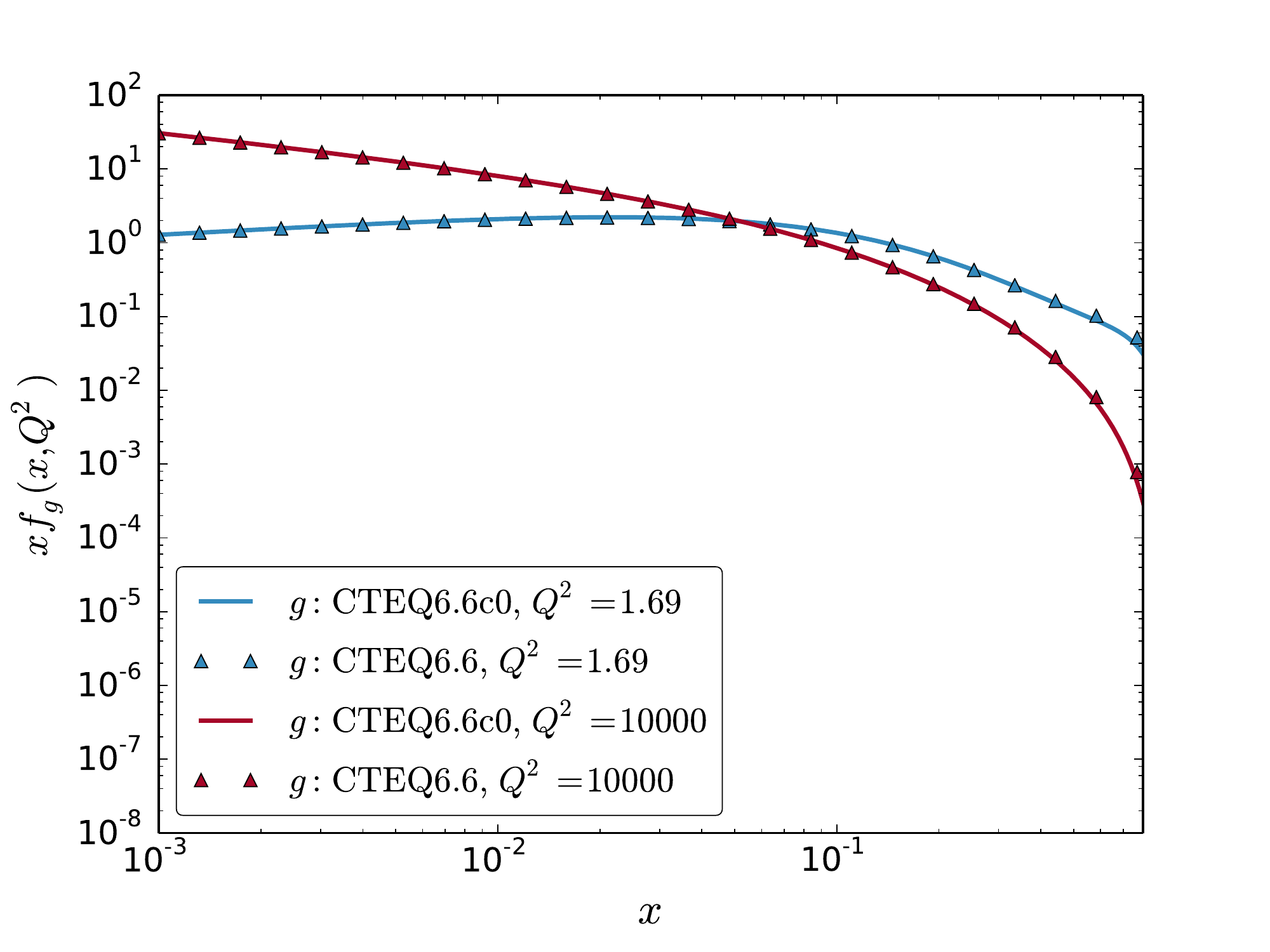}}
\subfigure[]{
\label{fig:gluonb}
\includegraphics[angle=0,width=0.48\textwidth]{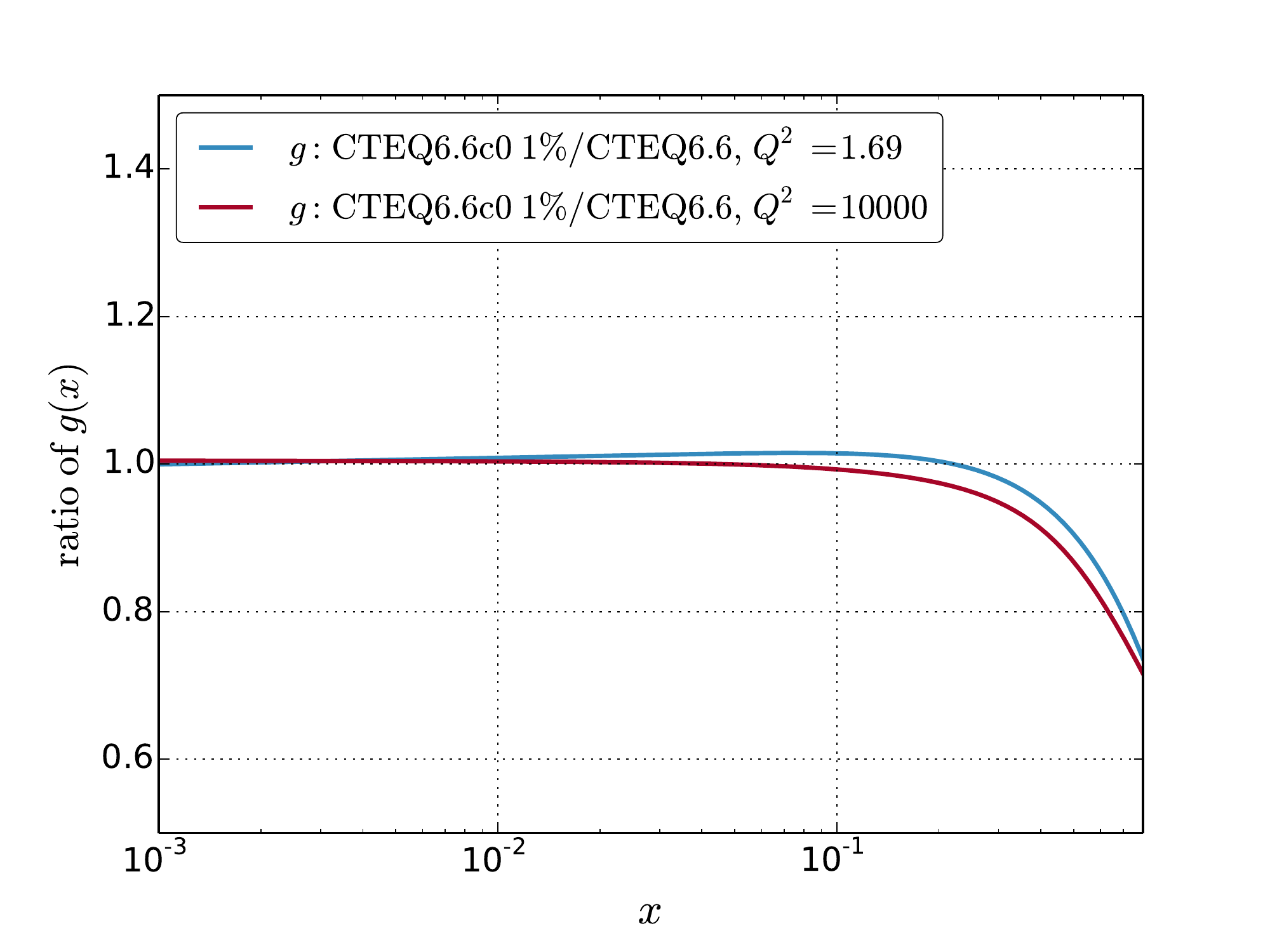}}
\caption{(a) Comparison of the CTEQ6.6c0 (solid line) and the CTEQ6.6 (triangles) gluon distributions.
(b) Ratio of the CTEQ6.6c0 and the CTEQ6.6 gluon distributions.
The results are shown as function of $x$ for two scales, $Q^2=1.69$ and $Q^2=10000\ \GeV^2$.
}
\label{fig:gluon}
\end{center}
\end{figure}
The inclusion of the intrinsic charm distribution will 
alter the other parton distributions, most notably the gluon PDF.
In order to gauge this effect we compare in Fig.~\ref{fig:gluon} the gluon distribution from the CTEQ6.6c0 analysis
with the one from the standard CTEQ6.6 fit.  
Fig.~\ref{fig:gluona} shows the $x$-dependence of the gluon distribution 
for two scales, $Q^2=1.69$ and $Q^2=10000\ \GeV^2$; Fig.~\ref{fig:gluonb} shows the ratio of these curves. 
For small $x$ ($x<0.1$) the gluon PDF is not affected by the presence of a BHPS-like intrinsic charm component
which is concentrated at large $x$.
At $x \simeq 0.7$, the CTEQ6.6c0 gluon is suppressed by about 20\% with respect to CTEQ6.6,
and this is relatively insensitive to the value of  $Q^2$. We note that at large-$x$, 
the gluon distribution is already quite small and the uncertainty of the gluon PDF is sizable
(of order of 40 -- 50\% for the CTEQ6.6 set).
The difference between the gluon distributions is {\em slightly} enhanced when evolving from the input scale $Q^2=1.69\ \GeV^2$
to the electroweak scale $Q^2=10000\ \GeV^2$, but it is still much smaller than the PDF uncertainty. 
We conclude that for most applications,  adding a standalone intrinsic charm distribution to an
existing standard  global analysis of PDFs is internally consistent and leads to only a small error.  
Moreover, for the case of intrinsic bottom which is additionally suppressed, 
the accuracy of the approximation will be even better. 

\begin{figure}
\begin{center}
\subfigure[]{
\label{fig:bca}
\includegraphics[angle=0,scale=0.35]{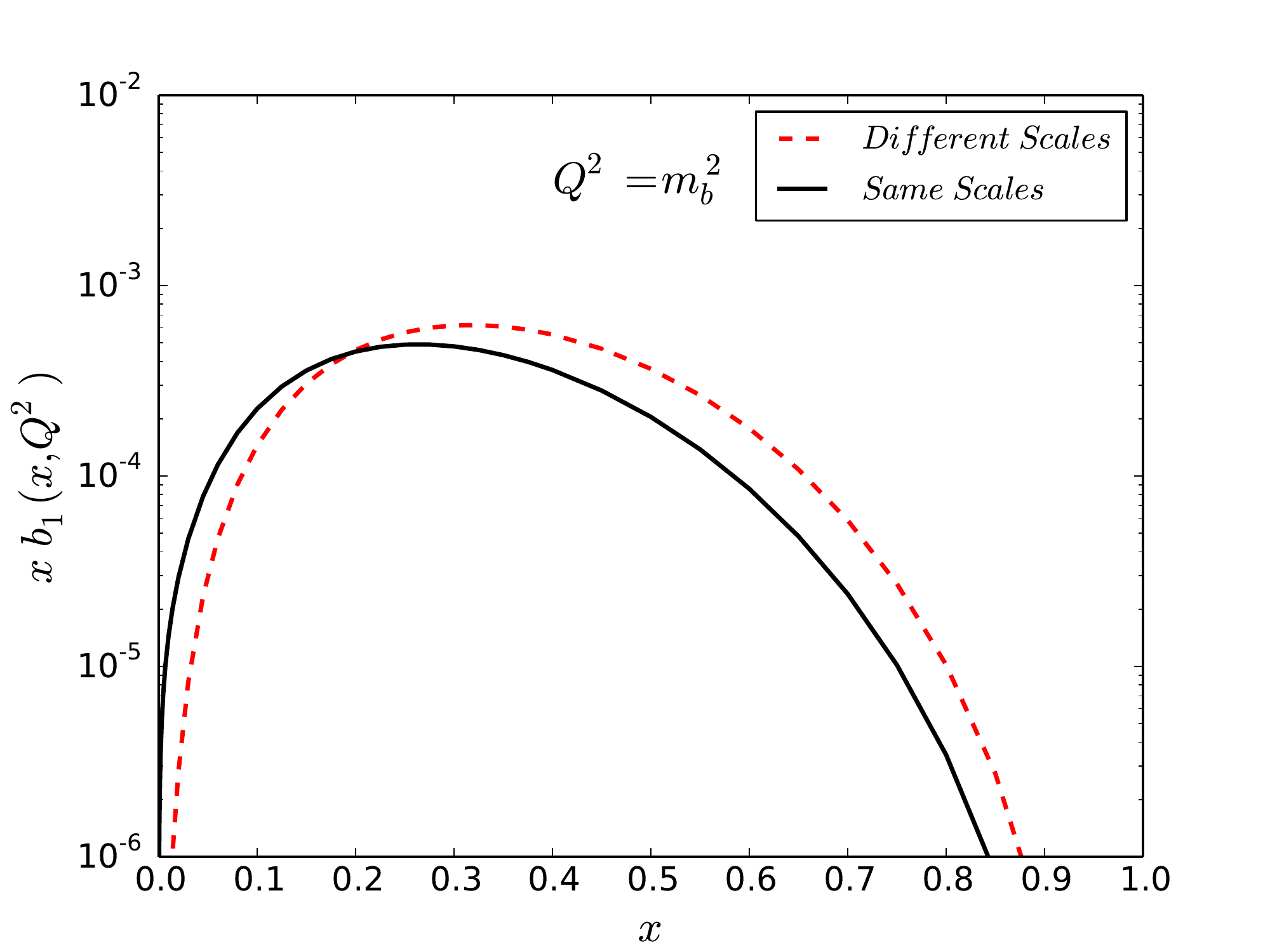}}
\subfigure[]{
\label{fig:bcb}
\includegraphics[angle=0,scale=0.35]{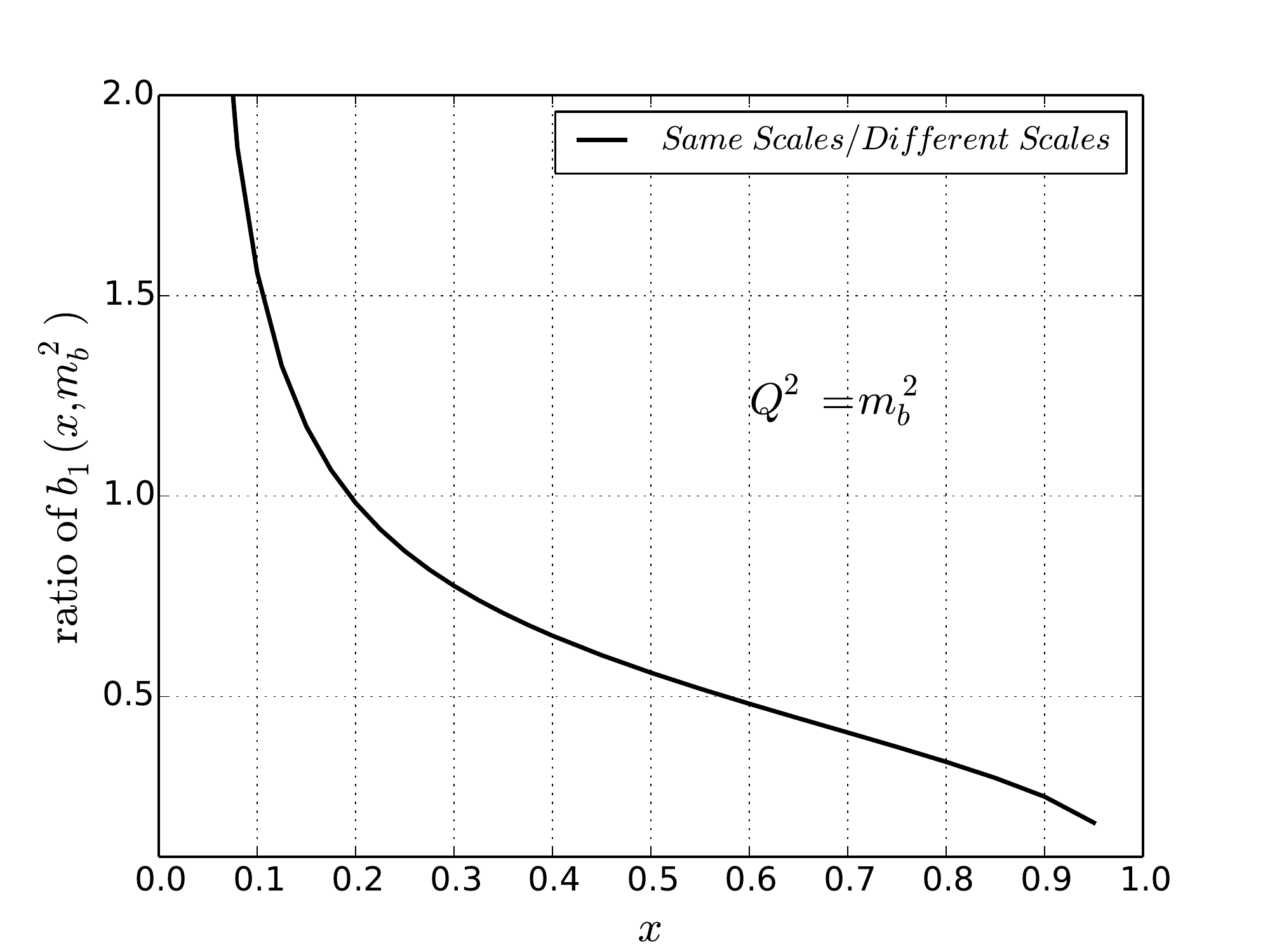}}
\caption{(a) Comparison of {\em Different Scales}, Eq.\ \eqref{eq:bc1},
and {\em Same Scales}, Eq.~\eqref{eq:bc2}, boundary conditions for the IB PDF at the scale
$Q=m_b=4.5\ \GeV$. Fig.~(b) shows the ratio of the two curves in (a).}
\label{fig:bc}
\end{center}
\end{figure}

Another source of uncertainty is the choice of boundary conditions
for the intrinsic distributions. 
For the IB PDF we have presented two equally compelling choices: 
the {\it Different Scales} ansatz of Eq.~\eqref{eq:bc1}
and 
the {\it Same Scales} one of Eq.~\eqref{eq:bc2}.
In Fig.~\ref{fig:bca} we compare the 
{\it Different Scales} (dashed line) and the {\it Same Scales} boundary conditions (solid line) at the scale $Q=m_b=4.5\ \GeV$.
As expected, the {\it Same Scales} boundary condition leads to a softer distribution due to the evolution from $m_c$ to $m_b$.
The ratio of these two distributions varies between $2$ at small $x$ and $\frac{1}{2}$ at large $x$
({\it cf.},  Fig.~\ref{fig:bcb}).%
    \footnote{We show here the ratio for $x\gtrsim0.1$;  at lower $x$ values the intrinsic distribution $b_1$
    is negligible compared to the perturbative $b_0$ component, as can be seen in Fig.~\ref{fig:ratio_Q0_Q1}.}
Note that the uncertainty due to the normalization is larger; for example, the ratio of the IC distributions $c_1(x,m_b)$ from 
CTEQ6.6c1 and CTEQ6.6c0 is about 3.5. Furthermore, there is also a freedom in the parametric $m_c^2/m_b^2$ factor
depending on which values are used for the heavy quark masses.
Therefore, in the following we use the {\it Same Scales} boundary condition for our numerical studies and consider the normalization as a free parameter.

For completeness we also provide similar validation using the parton--parton luminosities
\begin{equation}
\frac{d\mathcal{L}_{ij}}{d\tau}(\tau,\mu) = \frac{1}{1+\delta_{ij}}\int_\tau^1 \frac{dx}{x}
                              \Big[ f_i(x,\mu)f_j(\tau/x,\mu) + f_j(x,\mu)f_i(\tau/x,\mu) \Big],
\label{eq:pplumi}
\end{equation}
where $\tau=x_1 x_2$. 
We can consider the production of a heavy final state particle $H$ at the LHC ($pp\to H$) of mass $m_H^2=\hat{s}$,
where $\hat{s}=x_1 x_2 S = \tau S$ so that $\tau=m_H^2/S$.
This will allow us to estimate the effects of our approximation on the physical observables 
as a function of the mass scale $\sqrt{\tau}=m_H/\sqrt{S}$.
We discuss the relation of the parton--parton luminosities to the actual cross-section
in more detail in Sec.~\ref{subsec:lumi}.

\begin{figure}[!h]
\centering
\subfigure[]{
\label{fig:lumiRatioIC1a}
\includegraphics[width=0.48\textwidth]{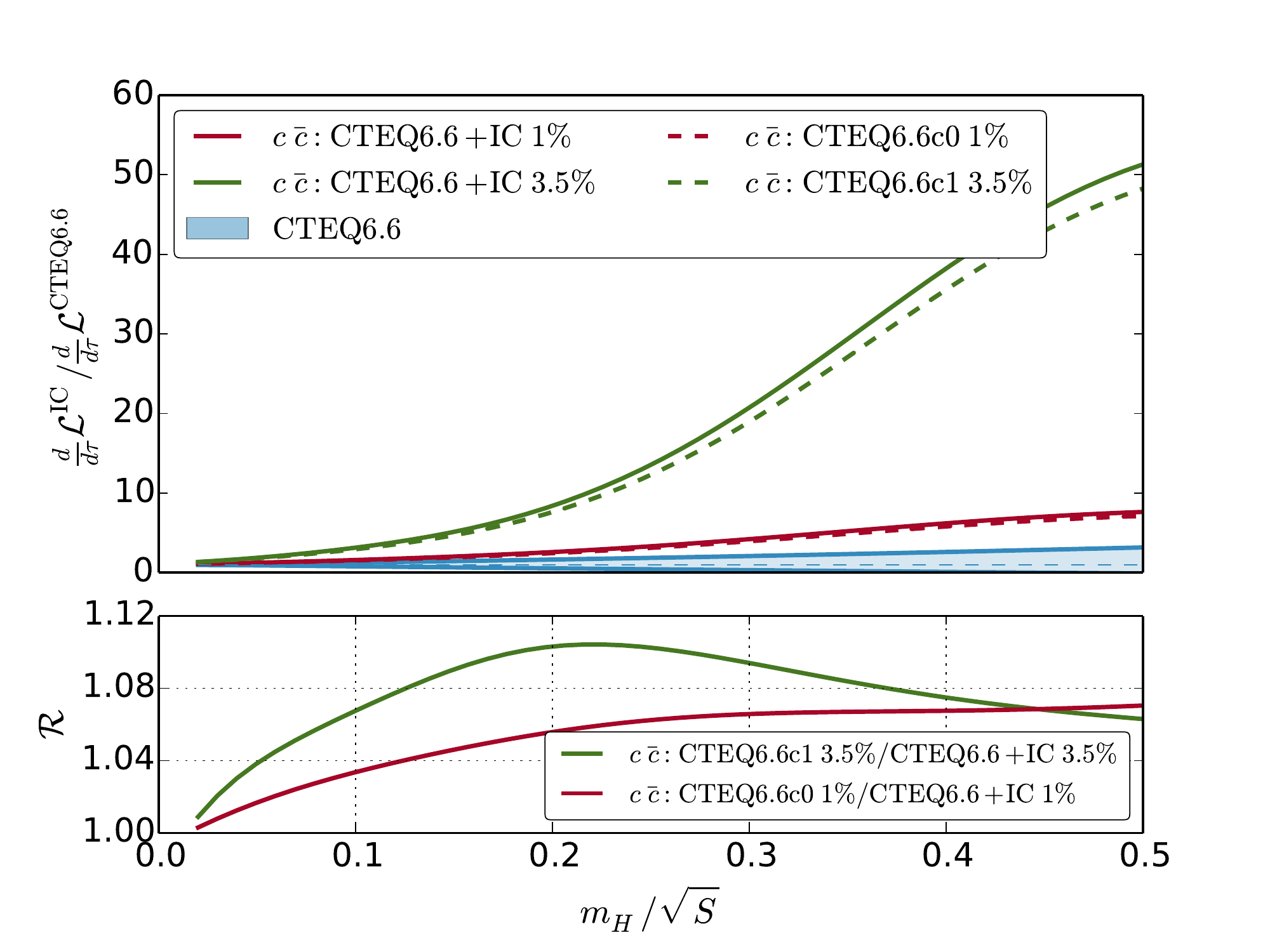}}
\subfigure[]{
\label{fig:lumiRatioIC1b}
\includegraphics[width=0.48\textwidth]{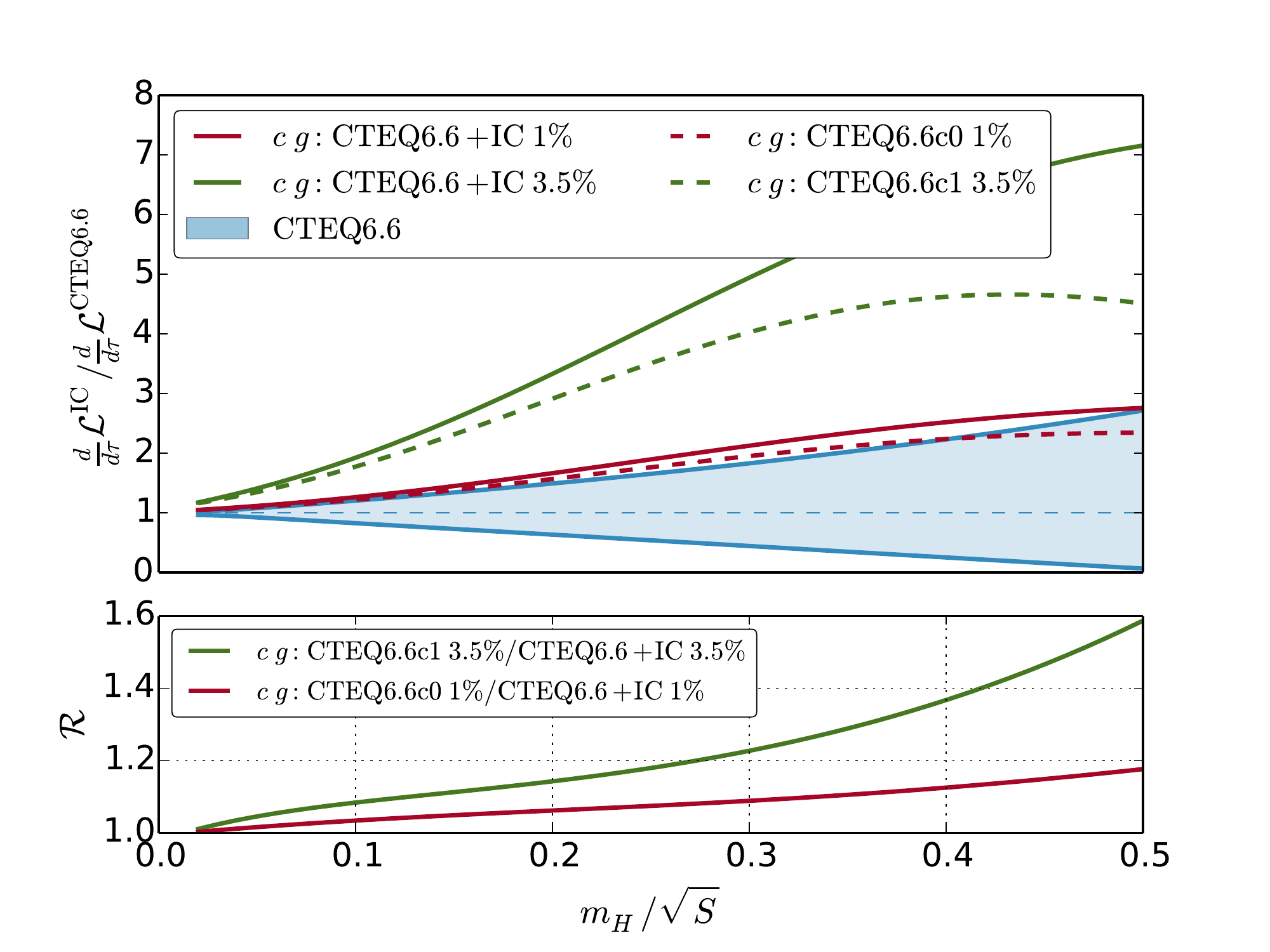}}
\caption{Parton--parton luminosities at the LHC with $\sqrt{S}=14$ TeV for (a) $\{c,\bar{c}\}$ 
and (b) $\{c,g\}$  calculated using:
our IC and CTEQ6.6 PDFs (solid lines),
CTEQ6.6c PDFs (dashed lines). 
For comparison, we show the PDF uncertainty (blue band) for the CTEQ6.6 PDFs.
The upper plots show ratio of the corresponding luminosities compared to the result obtained
with the central fit of CTEQ6.6. The lower plots present the ratio ${\cal R}$ of luminosities calculated using
our IC plus CTEQ6.6 PDFs compared to the result with CTEQ6.6c0 (CTEQ6.6c1) PDFs.
Note that the red curves represent cases with 1\% IC normalization and the green
ones cases with 3.5\% IC normalization.
}
\end{figure}

In Fig.~\ref{fig:lumiRatioIC1a} we show the ratio of $c \bar{c}$ luminosities for the IC 
with two choices of  normalizations; we compare the results  obtained with our approach 
to the CTEQ6.6c0 and CTEQ6.6c1 sets. 
We see that our error on the luminosities  (the difference between corresponding solid and dashed lines)
is smaller than 10\% across the full range of $\tau=m_H^2/S$ values.
Additionally, this error is much smaller than the
difference between CTEQ6.6 and scenarios with IC (distance between blue band and red/green lines).
Similarly, in 
Fig.~\ref{fig:lumiRatioIC1b} we show the ratio of luminosities for the $cg$ combination. 
In this case the error of our method is larger;  it is around 10\% for the IC 1\% normalization, 
and  it increases substantially for the case of IC 3.5\% normalization. Note that for the IC 3.5\% normalization, the deviation of our approximation from CTEQ6.6c1
is of the same order as the CTEQ6.6 PDF error band. 

Thus, we conclude our approach provides a good approximation
for  $c \bar{c}$ luminosities  for an IC with either 1\% or 3.5\% normalization. For $cg$ luminosities,  this approximation is quite reasonable 
for an IC with 1\% normalization; however, for  3.5\% normalization, it is only sufficient
to obtain a rough estimate of the effects. On the other hand, if the IC component is this large it 
should not be difficult to observe. 

Since the IB case can be obtained by scaling IC
with the  $m_c^2/m_b^2$ suppression factor, 
our approximation will work perfectly well 
for both the $b\bar b$ and the $b g$ luminosities due to the smaller normalization.

\section{Possible effects of IC/IB on LHC observables}
\label{sec:numerics}

In this section, we investigate the  effects of intrinsic heavy quarks on observables
at the LHC. We study the effects of both IC and IB on parton--parton luminosities
at 14 TeV LHC.
This allows us to assess the relevance
of a non-perturbative heavy quark component for the production of new heavy
particles coupling to the SM fermions.

\subsection{IB versus IC and prospects of observing IB}
\begin{figure}
\begin{center}
\includegraphics[width=0.5\textwidth]{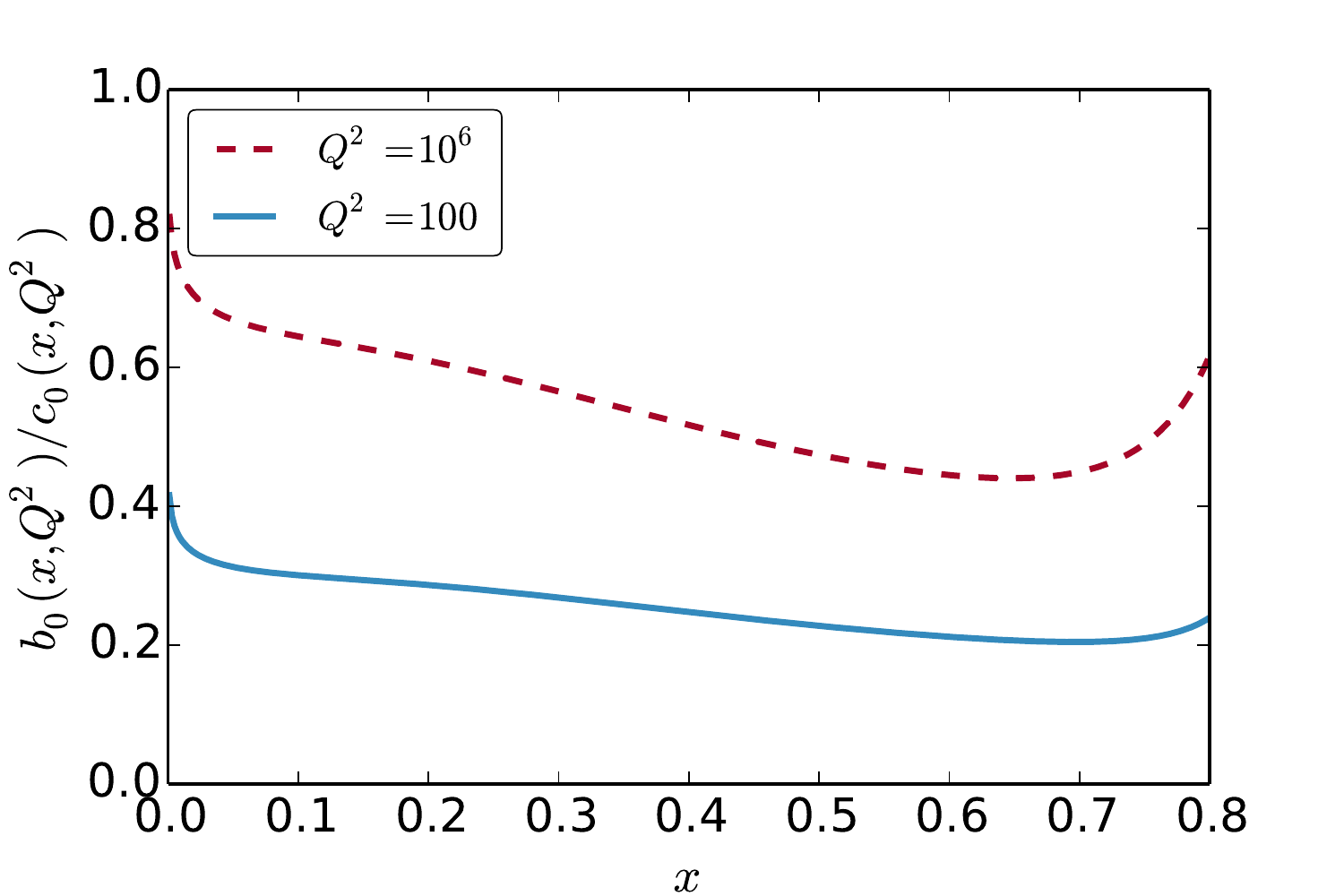}
\caption{
Ratio of the dynamically generated bottom and charm parton distributions at $Q^2=100$ and
$Q^2=10^6$ GeV$^2$ as a function of $x$.
The results for $x > 0.7$ are not reliable due to instabilities
in the CTEQ PDF grids for the charm distribution
(note however, that for that large $x$ values the perturbative PDFs
are nearly zero and are negligible).
}
\label{fig:ratio_b0_c0}
\end{center}
\end{figure}

\begin{figure}
\begin{center}
\subfigure[]{
\label{fig:c0c1}
\includegraphics[width=0.48\textwidth]{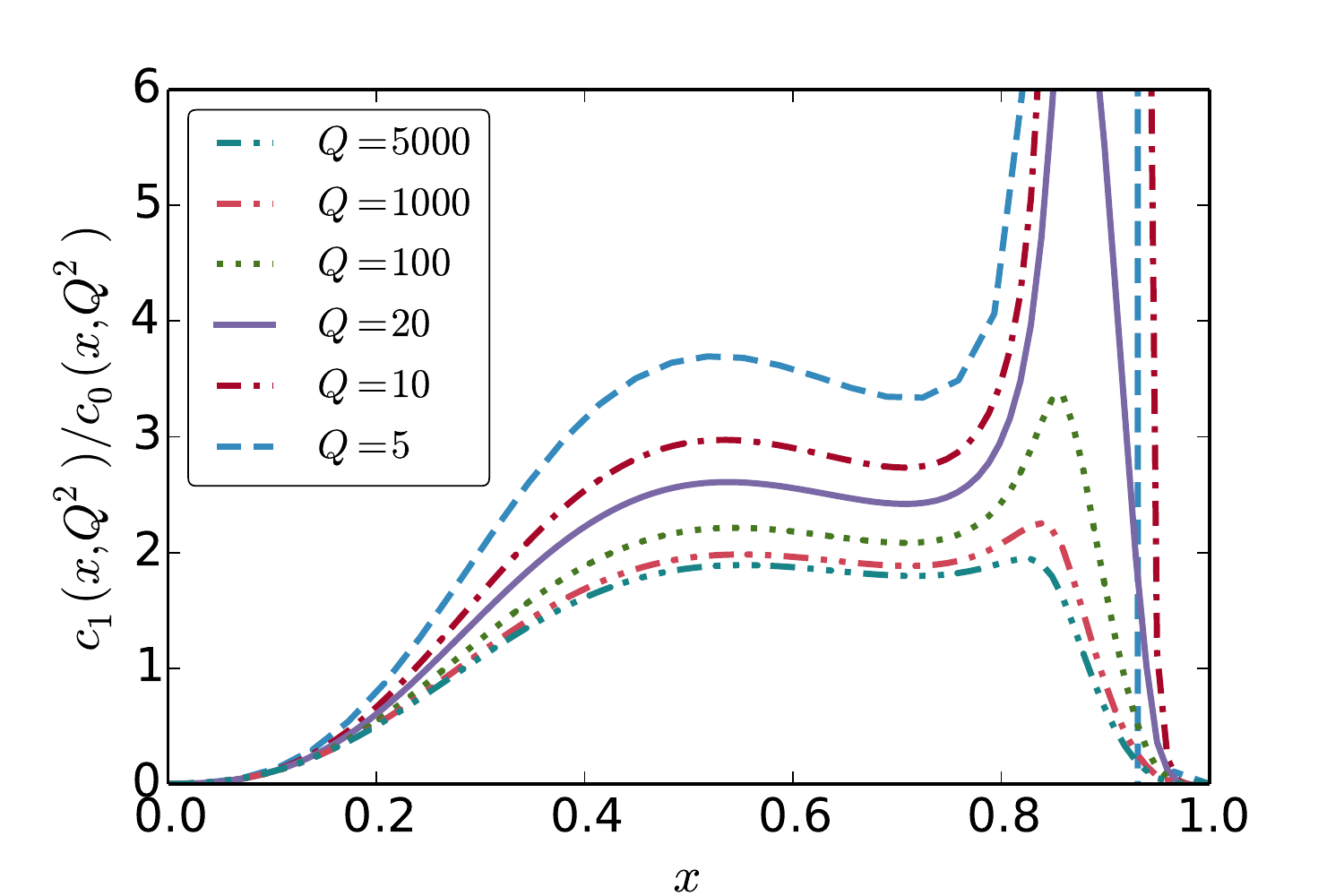}}
\subfigure[]{
\label{fig:b0b1}
\includegraphics[width=0.48\textwidth]{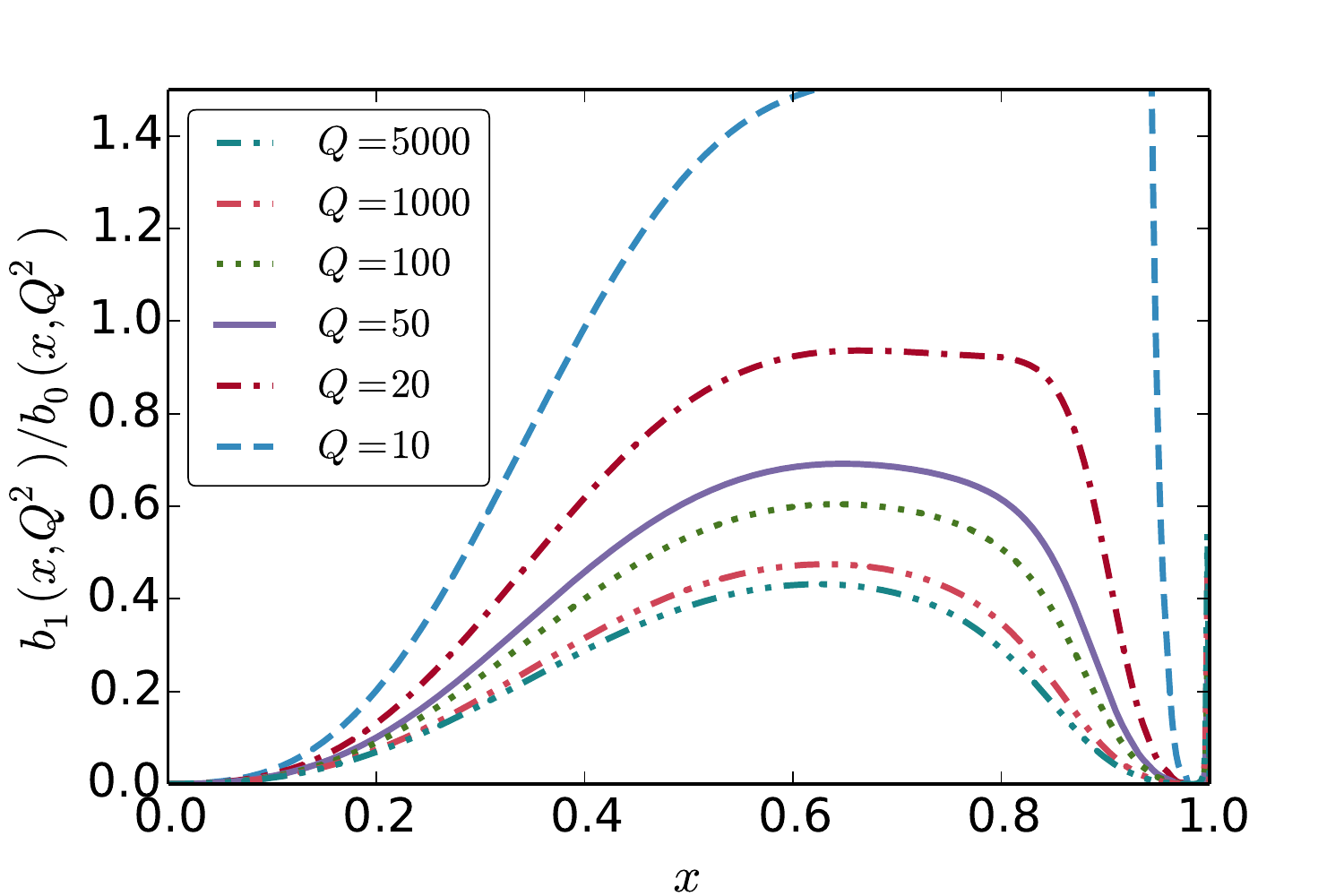}}
\caption{
Ratio of intrinsic and dynamically generated heavy quark PDFs for
(a) charm and (b) bottom quarks.
The distributions $c_1(x,Q^2)$ and $b_1(x,Q^2)$
have been generated using the  
non-singlet evolution equation \protect\eqref{eq:DGLAP4}
with the 
BHPS form \protect\eqref{eq:bhps} (properly normalized to 1\%) 
and the 
{\it Same Scales} boundary condition  \protect\eqref{eq:bc2}.
$c_0(c,Q^2)$ is the CTEQ6.6 charm distribution and
$b_0(x,Q^2)$ is the CTEQ6.6c0 bottom distribution.
As in Fig.~\ref{fig:ratio_b0_c0}, the results in (a) are not reliable for $x > 0.7$
due to
instabilities in the CTEQ6.6 grids for very low values of the $c_0$ distribution.
}
\label{fig:ratio_Q0_Q1}
\end{center}
\end{figure}

The IB PDF is parametrically suppressed by a factor $m_c^2/m_b^2\simeq 1/10$ compared to the IC PDF.
However, whenever particles have couplings to the SM fermions proportional to the fermion mass,
the  $m_c^2/m_b^2$ suppression will be compensated by a factor of $m_b^2$ in the coupling.
In addition, the radiatively generated (extrinsic) bottom PDF is typically
a factor of $\sim$1/5 to $\sim$1/2 smaller than the extrinsic charm PDF  as shown in  Fig.~\ref{fig:ratio_b0_c0}.
Combining these pieces together we find  $b_1/b_0 < c_1/c_0$ which means that the 
 possible effects due to the IB will be less
pronounced than the ones due to the IC. The ratios $c_1/c_0$ and $b_1/b_0$
are shown for several scales
in Figs.\ \ref{fig:c0c1} and \ref{fig:b0b1}.

These figures are useful to illustrate the impact of the intrinsic heavy quark distributions on 
the physical observables.
For this purpose, we define the ratio $\kappa_c=1+c_1/c_0$ and similarly $\kappa_b=1+b_1/b_0$
which measures the  relative deviation expected due to the  IC/IB components. 
For example, if the $b$-quark initiated subprocesses of an observable contributes a fraction $r_b$ to a cross section
(say $r_b = 80\%$), 
the observable will be enhanced by a factor $[r_b \times \kappa_b + (1-r_b)]$ where 
$\kappa_b$ is evaluated at the $x$-value relevant for the specific process.

We note that  $\kappa_c$ is still sizable at  $Q=100\ \GeV$ with $\kappa_c \simeq 3$ ($\kappa_c \simeq 8$) 
for the case of 1\% (3.5\%) normalization.%
    \footnote{Note that we show only the case of 1\% normalization, however, due to the scale-invariance,
    the  3.5\% normalization can be obtained by applying a multiplicative factor 3.5.}
In contrast, the IB content is much smaller at this scale with $\kappa_b \simeq 1.6$ ($\kappa_b \simeq 3.1$) for the
case of 1\% (3.5\%) normalization of the IC due
to the  $m_c^2/m_b^2$ factor. 
Therefore, we expect processes at the electroweak scale (or heavier scales) to be much less affected by the presence
of an intrinsic bottom component. Nevertheless, in cases where a process is dominated by the $b$-quark initiated subprocesses
and where the large $x$ region is probed (for example at large rapidities) an enhancement by a factor 1.6 (or even 3.1)
might be visible.
In general, processes probing lower scales ($Q < 20\ \GeV$) and large $x$ ($x\gtrsim 0.5$)
will be better suited to find/constrain the intrinsic bottom component of the nucleon.

\subsection{Parton--parton luminosities at the LHC} 
\label{subsec:lumi}
We now turn our attention to the parton--parton luminosities and study the impact of a
non-perturbative heavy quark component on these quantities for the LHC at 14 TeV.
Using the factorization theorem of QCD for hadronic cross sections, one can express
the inclusive cross section for the production of a heavy particle $H$ as follows: 
\begin{equation}
\sigma_{pp\rightarrow H+X} = \sum_{ij}  \int_\tau^1 \int_{\tau/x_1}^1 dx_1dx_2 f_i (x_1,\mu) f_j(x_2,\mu) \hat{\sigma}_{ij \rightarrow H}(\hat{s})\ ,
\label{eq:factorization}
\end{equation}
where $\tau=x_1 x_2 = m_H^2/S$, $S$ is the hadronic center of mass energy, and  $\hat{s}=x_1 x_2 S$ is its partonic
counterpart. $f_{i}(x,\mu)$ denotes the PDF of parton $i$ carrying momentum fraction
$x$ inside the proton. Finally, $\mu$ is the factorization scale which in the following is
identified with the partonic center of mass energy $\hat{s}=m_H^2$. Equation~\eqref{eq:factorization}
can be re-written in the form of a convolution of partonic cross-sections and parton--parton
luminosities \cite{Quigg:2009gg,Han:2014nja},
\begin{equation}
\sigma_{pp\rightarrow H+X} = \sum_{ij} \int_\tau^1 d\tau \ \displaystyle \frac{\mathcal{L}_{ij}}{d\tau}\ \hat{\sigma}_{ij}(\hat{s}),
\label{eq:lumidef}
\end{equation}
where $\frac{d\mathcal{L}_{ij}}{d\tau}(\tau,\mu)$ has been introduced in Eq.~\eqref{eq:pplumi}. 
\noindent
All the results of this section have been obtained using the CTEQ6.6 PDF set~\cite{Nadolsky:2008zw}
supplemented with our approximate IC and IB PDFs constructed using the procedure presented
in Sec.~\ref{sec:intrinsic}.

\begin{figure}[h]
\begin{center}
\subfigure[]{
\label{fig:luminositiesa}
\includegraphics[width=0.65\textwidth]{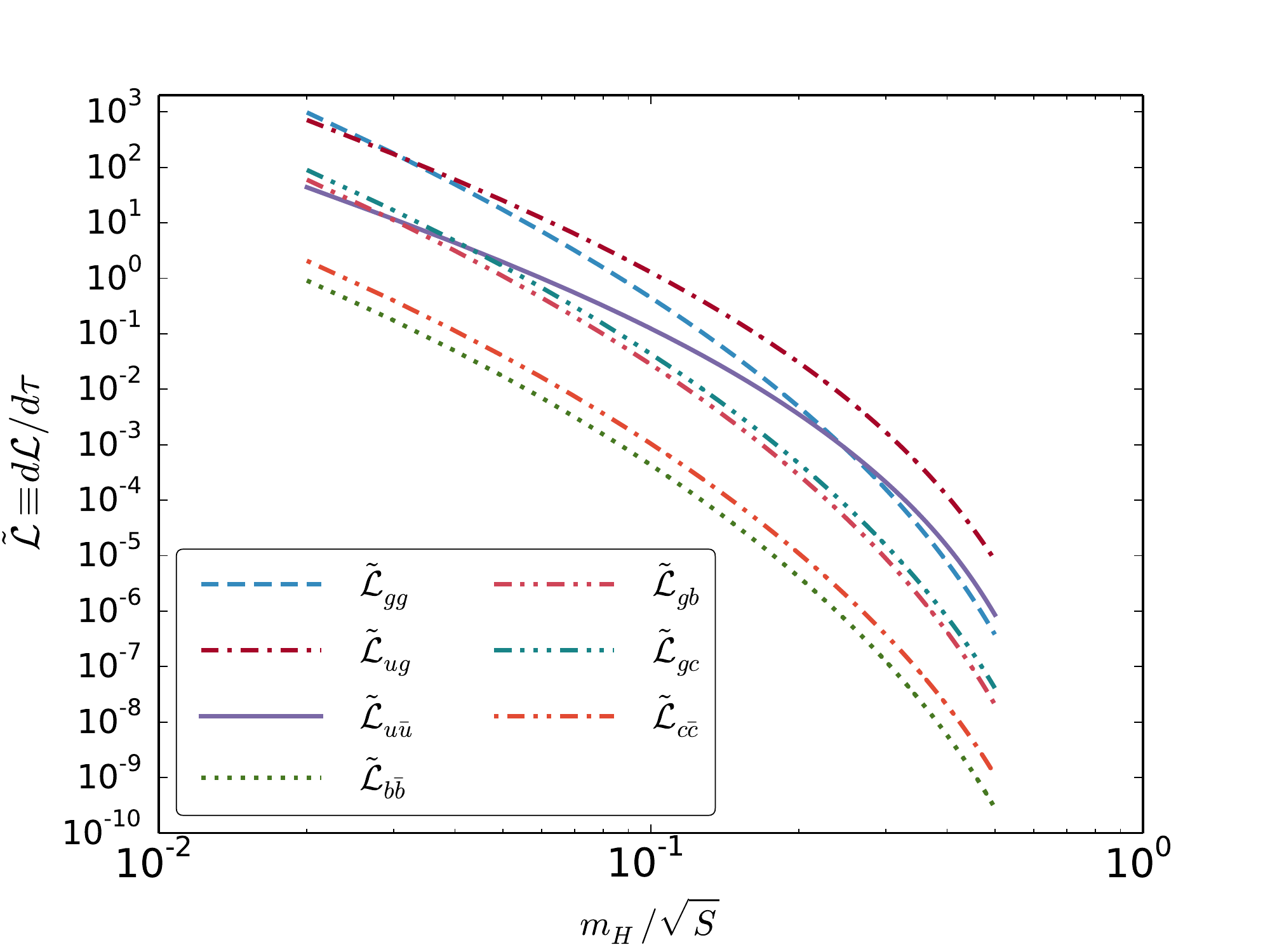}}
\subfigure[]{
\label{fig:luminositiesb}
\includegraphics[width=0.65\textwidth]{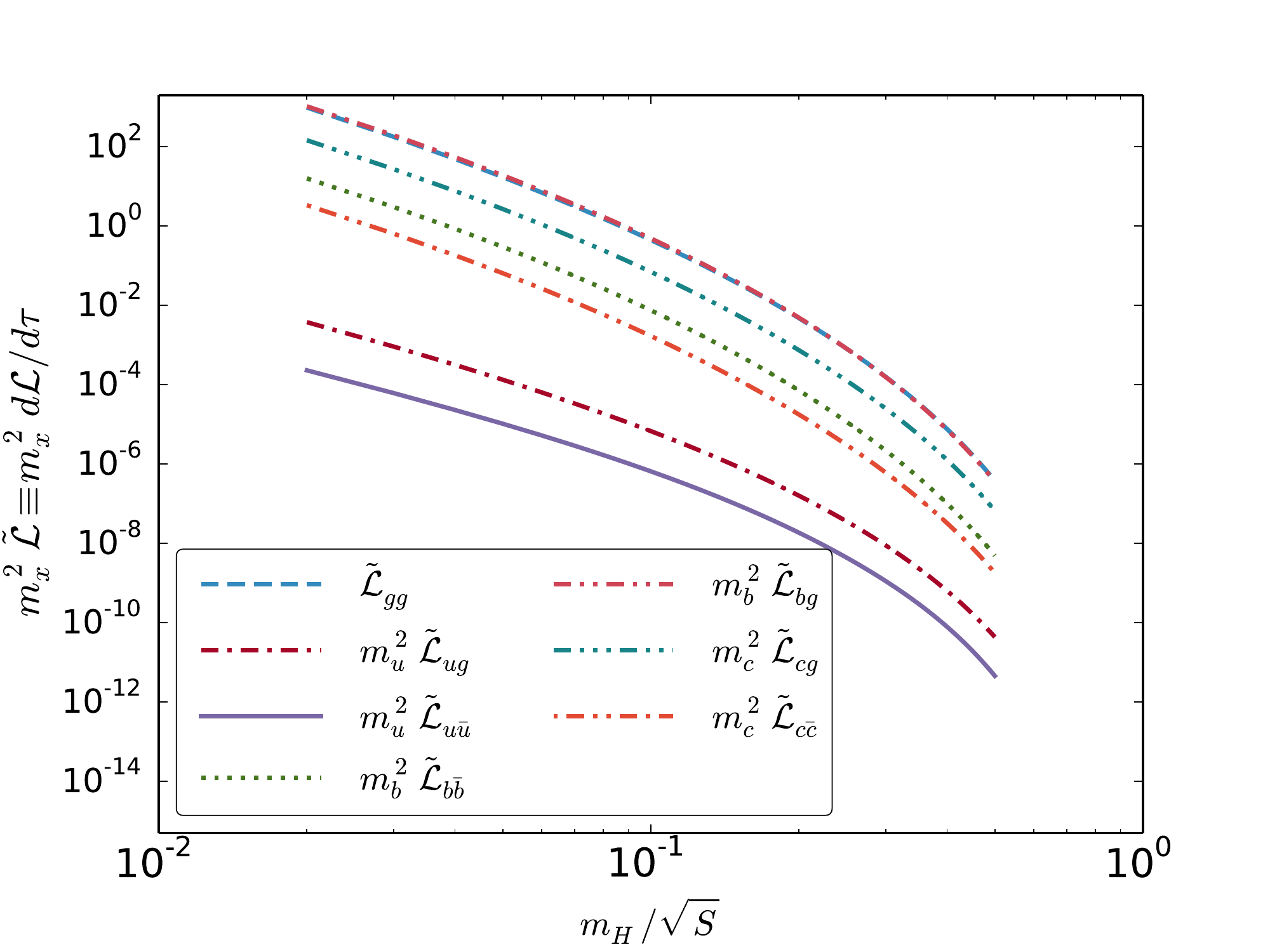}}
\caption{(a) Different parton--parton luminosities as a function of $\sqrt{\tau}=m_H/\sqrt{S}$
for the LHC14 calculated using CTEQ6.6 PDFs. 
For large $\tau$, the ordering of the curves is:
$ug \gg u \bar u > gg \gg gc > gb \gg c \bar c  > b \bar b$.
(b) Rescaled  
parton--parton luminosities ($m_i^2  d\mathcal{L}_{ij}/d\tau$) for the LHC14 calculated using CTEQ6.6 PDFs. 
For comparison, we also show the un-rescaled gluon--gluon luminosity.
For large $\tau$, the ordering of the curves is:
$gg \simeq gb > gc \gg b\bar b > c \bar c \gg u g \gg u \bar u$.
Note that by coincidence the gluon--gluon luminosity, $\mathcal{L}_{gg}$, agrees
at the 10\% level with the scaled $gb$ luminosity, $m_b^2 \mathcal{L}_{gb}$, so that the two curves
lie on top of each other in (b).
}
\end{center}
\end{figure}

In Fig.~\ref{fig:luminositiesa} we show different parton--parton luminosities, $d\mathcal{L}_{ij}/d\tau$,
for the LHC at 14 TeV (LHC14) as a function of $\sqrt{\tau}=m_H/\sqrt{S}$. We choose the range of $\sqrt{\tau}$
to be $[0.02,0.5]$ that corresponds to the production of a heavy particle of mass
$m_H\in[0.280,7]$ TeV which is roughly the range of values that will likely be probed at the LHC14.
As can be seen, at large $\sqrt{\tau}$, the parton--parton luminosities respect the following ordering:
$ug \gg u \bar u > gg \gg gc > gb \gg c \bar c  > b \bar b$.
Consequently, one can generally conclude that heavy quark initiated subprocesses play a minor role
in {\it most} processes where a heavy state is produced.

One exception would be SM extensions where the couplings to the first two generations
are suppressed or vanish so that the $gb$ or $b\bar b$ channels can dominate;
typically this is done in order to avoid experimental constraints from low energy
precision observables or flavor changing neutral currents. 
Of course, unless the couplings to the $gb$ or $b\bar b$ channels are enhanced, these scenarios
have tiny cross sections and will be difficult to measure at the LHC. 

However, if the couplings are enhanced by factors of the quark mass, 
the hierarchy of the contributions can change dramatically. 
This can happen when the heavy state has couplings to the
Standard Model particles proportional to their masses such as the SM Higgs or the Higgs particles in 2HDM models.
For example, in Fig.~\ref{fig:luminositiesa} we show the parton-parton luminosities 
with no enhancement factors; 
 in Fig.~\ref{fig:luminositiesb} we show the same but with additional factors proportional to the heavy quark mass;
the change is dramatic.
Taking the quark masses into account, the high $\tau$ region now exhibits the following hierarchy:
$gg \simeq gb > gc \gg b\bar b > c \bar c \gg u g \gg u \bar u$.
In this case the heavy quark initiated subprocesses could play the dominant role,
apart from the $gg$ initiated subprocesses which would contribute via an 
effective, model-dependent, heavy quark loop-induced $ggH$ coupling.

\begin{figure}[!h]
\centering
\includegraphics[width=\textwidth]{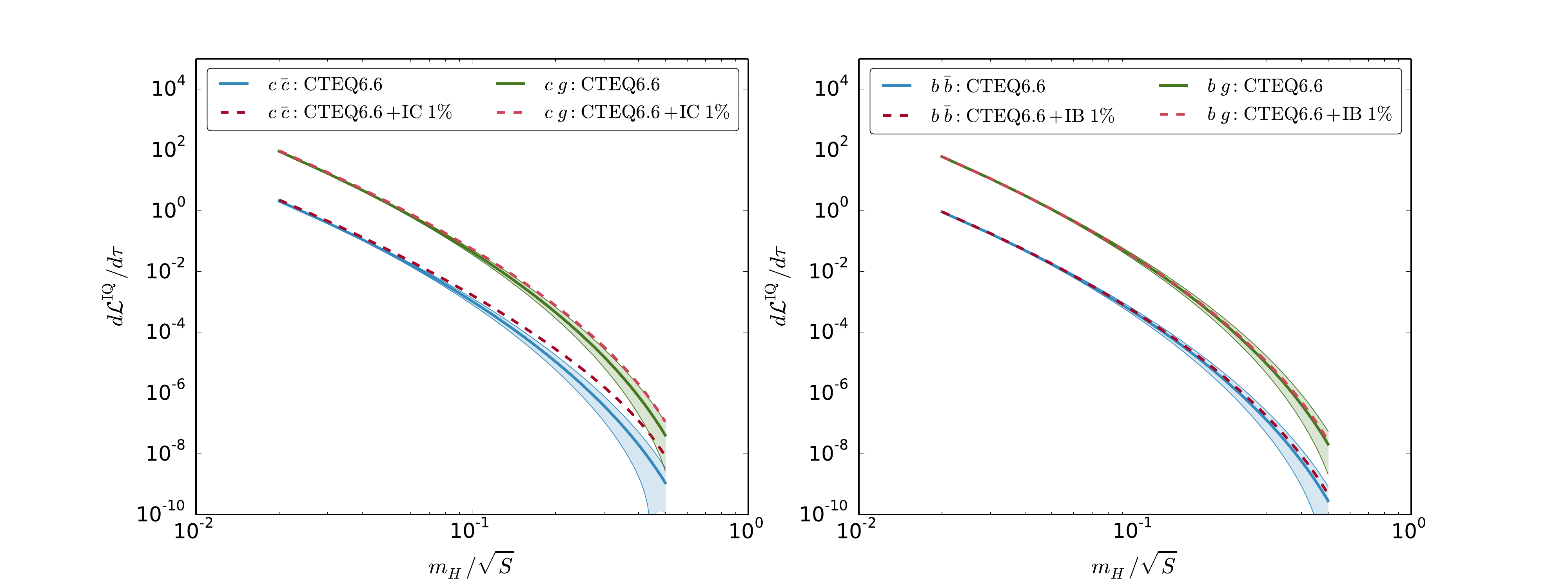}
\caption{Parton--parton luminosities 
$d\mathcal{L}_{c g}/d\tau$, $d\mathcal{L}_{c \bar c}/d\tau$ (left) and
$d\mathcal{L}_{b g}/d\tau$, $d\mathcal{L}_{b \bar b}/d\tau$ (right)
at the LHC14 as a function of $\sqrt{\tau}=m_H/\sqrt{S}$.
Shown are results without an intrinsic heavy quark component using the CTEQ6.6 PDFs (solid lines)
including the band due to the PDF uncertainties and the same quantities calculated with addition of IC normalized to 1\%
and the corresponding IB component (dashed lines). 
}
\label{fig:lumiwerrors}
\end{figure}

To explore how the presence of IC and IB would 
affect physics observables  with a non-negligible heavy quark initiated subprocesses, 
in Fig.~\ref{fig:lumiwerrors}, we present the parton--parton luminosities 
	$d\mathcal{L}/d\tau$, $d\mathcal{L}/d\tau$ with and without the intrinsic components. 
While the impact of the IC component for $\sqrt{\tau}>0.1$ is clearly visible, the
corresponding effect for the bottom-quark is smaller and lies inside the PDF uncertainty band.
Generally, the enhancement is larger for the processes initiated by two heavy quarks $\{c\bar{c},b\bar{b}\}$  since the intrinsic component is then 
``squared" in the luminosities.

\begin{figure}[t]
	\centering
	\includegraphics[width=\textwidth]{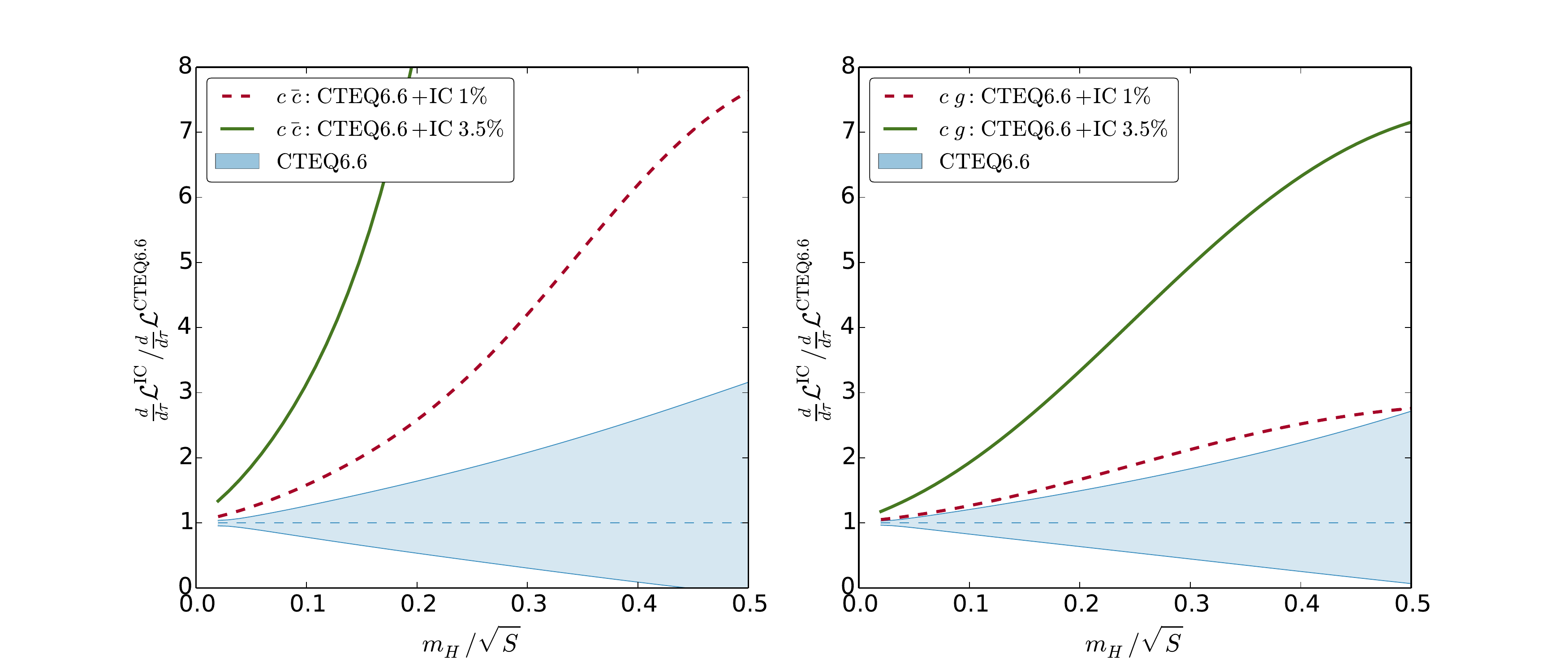}
	\caption{Ratio of $c \bar c$ luminosities (left) and $c g$ luminosities (right) at the LHC14 
	for charm-quark PDF sets with and without an intrinsic component as a function of $\sqrt{\tau}=m_H/\sqrt{S}$. 
	The ratio for the $c \bar c$ luminosity (solid, green line) in the left figure reaches values of 50 at $\sqrt{\tau}=0.5$. In addition to the curves with 1\% normalization (red, dashed lines) we include the results for the 3.5\% normalization (green, solid lines)
which was found to be still compatible with the current data~\cite{Nadolsky:2008zw}.
	}
	\label{fig:ratio_w_uncertainty_c}
\end{figure}

\begin{figure}[!h]
\centering
\includegraphics[width=\textwidth]{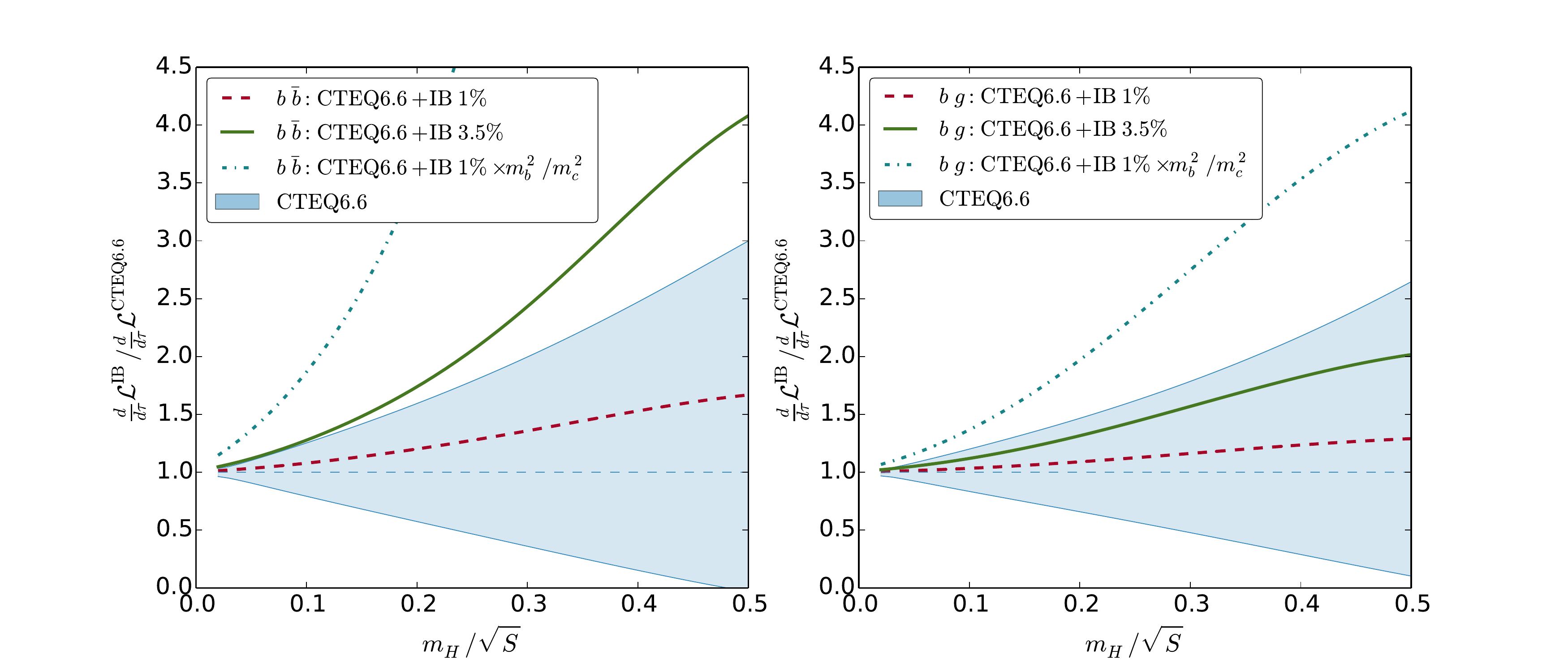}
\caption{Ratio of luminosities at the LHC14 for bottom-quark PDF sets with different
normalizations of the intrinsic bottom component. The plot has been truncated, and the $b\bar{b}$
luminosity in the extreme scenario reaches about 17 at $\sqrt{\tau}=0.5$.}
\label{fig:ratio_w_uncertainty_b}
\end{figure}

In order to precisely quantify 
the impact of the intrinsic components
in Figs.~\ref{fig:ratio_w_uncertainty_c} and~\ref{fig:ratio_w_uncertainty_b} we show
the ratios of luminosities for charm and bottom with and without an intrinsic contribution
for 1\% and 3.5\% normalizations.
Furthermore, since there are no experimental constraints on the IB normalization,
in Fig.~\ref{fig:ratio_w_uncertainty_b} we also include an extreme scenario where 
we remove the usual  $m_c^2/m_b^2$ factor; thus, the first moment of the IB is 1\% at the initial scale $m_c$.

For the 1\% normalization the  $c\bar c$ luminosity ratio 
grows as large as 7 or 8 for  $\sqrt{\tau}=0.5$, 
and for a 3.5\% normalization it
becomes extremely large and reaches values of up to 50.
From these figures we can clearly see that the effect of the 3.5\% IC is substantial
and can affect observables sensitive to $c\bar{c}$ and $cg$ channels. As expected, in the case of IB the effect is smaller but for the $b \bar{b}$ luminosity  the
IB with 3.5\% normalization leads to a curve which lies clearly above 
the error band of the purely perturbative result. 
In the extreme scenario (which is not likely but by no means excluded) the IB component has a big effect 
on both the $b\bar{b}$ and $bg$ channels.

\section{Discussion}
\label{sec:discussion}

We have demonstrated that the scale evolution of intrinsic heavy quark distributions (both charm and bottom)
is governed by a non-singlet evolution equation to a very good approximation.
Furthermore, the small intrinsic heavy quark distribution does not significantly influence  the
other parton distributions or the sum rules of a global analysis. This observation holds to a very good precision
for the intrinsic bottom case,  but also works reasonably well for the intrinsic charm case (if the momentum fraction is not too large).
Therefore, it is possible to perform a standalone analysis of the intrinsic heavy quark distribution
and to combine it with the PDFs of a standard global analysis with dynamically generated heavy quark 
distributions\footnote{Needless to say, that the intrinsic heavy quark distribution could also be used together 
with the PDFs in a fixed-flavor-number scheme where no dynamically generated heavy quark distribution is present.}.
Note, this allows us to use any general PDF set and generate a matched IC or IB component 
without a global fit re-analysis. 

Based on this observation we have modeled an intrinsic bottom distribution 
and discussed its effect on the relevant parton--parton luminosities at the LHC14. 
As a general rule, the effects of IB are less pronounced than the ones from IC
due to the expected $m_c^2/m_b^2 \sim 0.1$ suppression factor. For example, we see from  Fig.~\ref{fig:ratio_Q0_Q1} (1\% normalization) that whereas 
$\kappa_c=1+c_1/c_0 \sim 3$ at scales $Q\sim 100$~GeV  and $x \in [0.4,0.8]$
the corresponding factor for IB is relatively small: 
$\kappa_b=1+b_1/b_0 \lesssim 1.4$. For a  3.5\% normalization, the curves in Fig.~\ref{fig:ratio_Q0_Q1} would be scaled
by a factor 3.5, such that $\kappa_b \lesssim 2.4$.

We then turned to a discussion of the parton--parton luminosities where the main results can be seen 
in Figs.\ \ref{fig:ratio_w_uncertainty_c} and \ref{fig:ratio_w_uncertainty_b}.

The $c\bar c$ luminosities are strongly enhanced for the IC with both 1\% and 3.5\% normalization as the 
heavy quark factors enter quadratically; 
but the effect is smaller for the $cg$ luminosity as there is only one factor of the heavy quark PDF. 
However, the effect is still significant; the  1\% IC lies at the edge of the PDF uncertainty band,
and the 3.5\% IC yields a factor of $\sim$5.

As expected, the enhancement of the $b\bar b$ luminosity is much smaller compared to the $c\bar c$ case. For the 1\% IB the curve lies well within the PDF uncertainty band. For the 3.5\% IB the results lies above this uncertainty band and predicts an enhancement of a factor 4 at $\sqrt{\tau}=0.5$.
For the $bg$ luminosity, both the 1\% IB and the 3.5\% IB curves lie within the PDF uncertainty band.
This band is largely driven by the uncertainty of the gluon distribution and might shrink in the future
such that the enhancement due to the IB could become significant.
For illustration, we have also included results for the extreme assumption that the probability
(first moment) for IB is 1\%.
In this case the enhancement is sizable and reaches a factor 17 for the $b\bar b$ case and a factor 4 for the
$bg$ case at $\sqrt{\tau}=0.5$.

In view of this, we can address the impact of IB on heavy new physics and certain electroweak processes
where the $bg$ or the $b\bar b$ channel plays an important role. For the 1\% and 3.5\% IB the enhancement for the $bg$-initiated subprocesses would be hidden within the PDF uncertainties.
The $b\bar b$-initiated subprocesses could be significantly enhanced in the case of 3.5\% IB.
The effect would be more pronounced for heavier states where however, the $b\bar b$ luminosity is extremely small
so that a measurement would be limited by statistics even for models with enhanced couplings to the $b$ quark.
All in all, we conclude that the IB will have limited impact on searches for heavy new physics at the LHC.
%
\section{Conclusions}
\label{sec:conclusions}

In this article, we presented a method to generate a matched IC/IB distributions
for any PDF set without the need for a complete global re-analysis. This allows one to easily carry
out a consistent analysis including intrinsic heavy quark effects. 
Because the evolution equation for the intrinsic heavy quarks decouples, 
we can freely adjust  the normalization of the IC/IB PDFs.

For the IB, our approximation holds to a very good precision.
For the IC, the error increases (because the IC increases), yet our method is still useful.
For an IC normalization of  1-2\%, the error is less than the  PDF uncertainties at the large-$x$ where the IC is relevant.
For a larger normalization, although the error may be the same order as the  PDF uncertainties, 
the IC effects also grow and can be separately distinguished from the case without IC. 
In any case, the IC/IB represents a non-perturbative systematic effect 
which should be taken into account.

The method presented here greatly simplifies our ability to search for, and place constraints upon, 
intrinsic charm and bottom compoments of the nucleon. This technique will facilitate
more precise predictions which may be  observed at future facilities such as an Electron Ion Collider (EIC), 
the Large Hadron-Electron collider (LHeC),  or AFTER\at LHC.

The PDF sets for intrinsic charm and intrinsic bottom discussed in this analysis
(1\% IC, 3.5\% IC, 1\% IB, 3.5\% IB) are available from the authors upon request.

\section*{Acknowledgment} 
We are grateful to T.\ Stavreva for her participation in the earlier stages of this project and to
S.\ Brodsky for useful discussions on the BHPS model.
This work was supported by the CNRS through a PICS research grant and the
Th\'eorie-LHC-France initiative.

\clearpage

\bibliographystyle{JHEP}
\bibliography{iQ} 

\end{document}